\documentclass{aa}

\usepackage{natbib}
\usepackage{threeparttable}
\usepackage{color,graphicx,dcolumn}
\usepackage{adjustbox,booktabs}
\usepackage[version=4]{mhchem}
\usepackage{amsmath,amssymb,latexsym}
\usepackage[varg]{txfonts}

\newcolumntype{d}{D{.}{.}{3.2}}

\begin{document}

\title{Photoinduced polycyclic aromatic hydrocarbon dehydrogenation:}
\subtitle{The competition between H- and \ce{H2}-loss}
 
\author{P.~Castellanos\inst{1,2} \and A.~Candian\inst{1} \and J.~Zhen\inst{3,4} \and H.~Linnartz\inst{2} \and A.~G.~G.~M.~Tielens\inst{1}}

\date{Received 12 April, 2018 / Accepted 02 June, 2018}

\titlerunning{PAH dehydrogenation}
\authorrunning{Castellanos et al.}

\institute{Leiden Observatory, Leiden University, P.O. Box 9513, NL-2300 RA Leiden, The Netherlands\\ \email{pablo@strw.leidenuniv.nl} \and Sackler Laboratory for Astrophysics, Leiden Observatory, Leiden University, P.O. Box 9513, NL-2300 RA Leiden, The Netherlands \and CAS Key Laboratory for Research in Galaxies and Cosmology, Department of Astronomy, University of Science and Technology of China, Hefei 230026, People's Republic of China \and School of Astronomy and Space Science, University of Science and Technology of China, Hefei 230026, People's Republic of China}

\abstract{Polycyclic aromatic hydrocarbons (PAHs) constitute a major component of the interstellar medium carbon budget, locking up to 10--20\% of the elemental carbon. Sequential fragmentation induced by energetic photons leads to the formation of new species, including fullerenes. However, the exact chemical routes involved in this process remain largely unexplored. In this work, we focus on the first photofragmentation steps, which involve the dehydrogenation of these molecules. For this, we consider a multidisciplinary approach, taking into account the results from experiments, density functional theory (DFT) calculations, and modeling using dedicated Monte-Carlo simulations. By considering the simplest isomerization pathways --- i.e., hydrogen roaming along the edges of the molecule --- we are able to characterize the most likely photodissociation pathways for the molecules studied here. These comprise nine PAHs with clearly different structural properties. The formation of aliphatic-like side groups is found to be critical in the first fragmentation step and, furthermore, sets the balance of the competition between H- and \ce{H2}-loss. We show that the presence of trio hydrogens, especially in combination with bay regions in small PAHs plays an important part in the experimentally established variations in the odd-to-even H-atom loss ratios. In addition, we find that, as PAH size increases, \ce{H2} formation becomes dominant, and sequential hydrogen loss only plays a marginal role. We also find disagreements between experiments and calculations for large, solo containing PAHs, which need to be accounted for. In order to match theoretical and experimental results, we have modified the energy barriers and restricted the H-hopping to tertiary atoms. The formation of \ce{H2} in large PAHs upon irradiation appears to be the dominant fragmentation channel, suggesting an efficient formation path for molecular hydrogen in photodissociation regions (PDRs).}


\maketitle

\section{Introduction}
\label{sec:intro}

Polycyclic aromatic hydrocarbons (PAHs) are generally accepted as carriers of the aromatic infrared bands (AIBs), a family of emission features that dominate the mid-infrared spectrum of most astronomical objects containing dust and  gas \citep[and reference therein]{Tielens2013}. Infrared space telescopes such as \textit{Spitzer} have allowed further insight into the composition of the astronomical PAH family and other large C-containing molecules, such as the fullerene \ce{C60}, which has been unambiguously identified in planetary nebulae \citep[e.g.,][]{Cami2010,gar12,ots13} and in photodissociation regions \citep[PDRs,][]{sell10,ber12,boe12,cas14}. \citet{ber12} noted a clear anticorrelation of the intensity of the PAH and fullerene IR-bands in NGC~7023. Based on this, they proposed that fullerenes can be formed in PDRs through UV photodestruction of large PAHs ($N_{\mathrm{C}} \geq 60$). The first step of this process involves the stripping of the hydrogen atoms from the edge of the molecule. This would be followed by \ce{C2}-loss and isomerization, effectively folding the dehydrogenated PAH into a closed fullerene structure \citep{ber15}. Support for this model has been provided by observations of other fullerene containing PDRs \citep{cas14}.

Recently, \citet{zhe14b} have presented experimental evidence supporting fullerene formation through PAH photodestruction. However, the details of this mechanism are still not understood. Studies regarding the first stage of such process --- dehydrogenation --- have been limited mostly to small PAHs and consider only the lower energy channels \citep{Ling95,Ling98,eke98,job14}. Appearance energies (AE) for the lowest energy channels, namely  H-, \ce{H2}- and \ce{C2H2}-loss, have been measured experimentally for PAHs up to coronene  \citep{joc94}. These experiments show that the preferred fragmentation channel is H-loss for all but one of the molecules in their study and that the AEs for H- and \ce{C2H2}-loss are very close in energy --- e.g., for the triphenylene cation (\ce{C18H12+}) they correspond to 16.1$\pm$0.3 and 16.7$\pm$0.3~eV, respectively \citep{joc94}.

\begin{figure*}[t!]
  \centering
  \includegraphics[width=\textwidth]{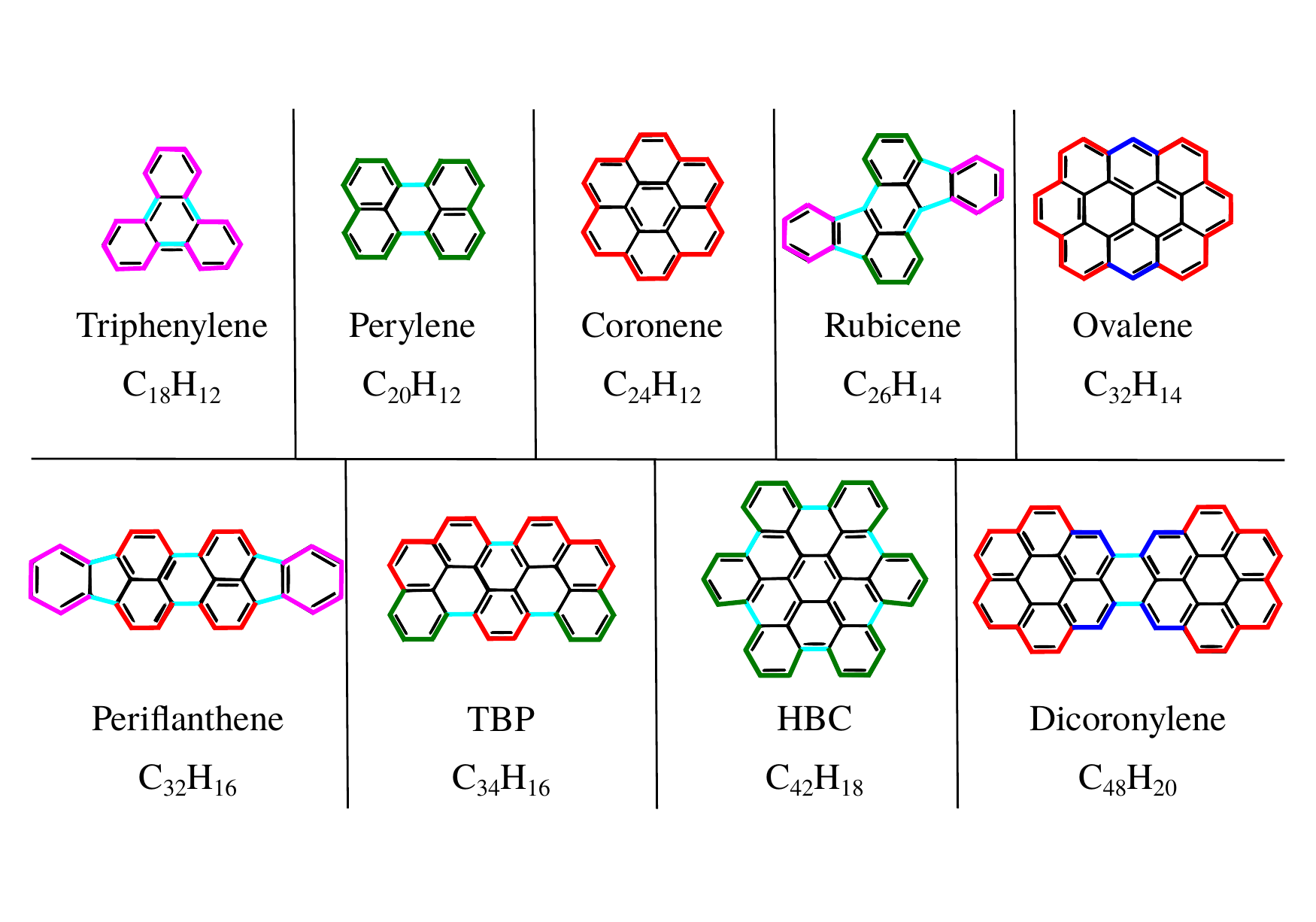}
  \caption{Structures and chemical formula of the nine PAHs studied in this paper. TBP identifies tribenzoperopyrene and HBC hexabenzocoronene, respectively. The different edge structures are highlighted with different colors: solo (blue), duo (red), trio (green) and quarto (fuchsia). Edge bonds without hydrogens attached correspond to bay regions (cyan), which can take different forms. \label{fig:pahs}}
\end{figure*}

All experimental studies on PAH dehydrogenation show the same pattern: photoproducts with an even number of hydrogen atoms have systematically higher abundances than those that retain an odd number of hydrogens \citep{eke98,job14,zhe14a,zhe14b}. Additionally, PAHs with less than 24 carbon atoms do not reach full dehydrogenation before acetylene (\ce{C2H2}) loss starts competing with pure dehydrogenation \citep{eke97,job14}, while larger PAHs appear to reach full dehydrogenation before carbon losses become important \citep{eke98,zhe14b}.

Direct fragmentation of PAHs is not the only process that needs to be taken into account upon irradiation. Isomerization in general and hydrogen hopping in particular, is known to proceed efficiently at internal energies below the threshold for direct dissociation \citep{bau14,Paris14,Chen15,tri17a,tri17b}. Considering a molecule with two adjacent hydrogen atoms on the edge as coronene for example, when a hydrogen shifts, the molecule hybridization  changes locally from $sp^2$ to $sp^3$, resulting in an aliphatic-like bond and an empty carbon center. Isomerization can affect the fragmentation behavior. For example, aliphatic C--H bonds are notoriously weaker than aromatic  C--H bonds \citep{Paris14,tri17a}. Additionally, there is the possibility to lose molecular hydrogen  from such an aliphatic site with a transition state significantly lower than the corresponding transition state from a fully aromatic structure \citep{Paris14}. This mechanism could be relevant as alternative formation route of \ce{H2} on PAH molecules in PDRs \citep[see][for a recent review]{Wakelam2017}. Other isomerization pathways lead to alterations of the carbon skeleton, i.e., the formation of dangling ethynyl or vinyl groups, or rearrangement of the hexagons into pentagons and/or heptagons \citep{bou16,tri17b}.

Here we report a systematic study of the H-loss pathway in PAHs under laser irradiation using ion trap time-of-flight mass spectrometry \citep{zhe14a} (Sect.~\ref{sec:experiment}). To interpret the experimental data, Density functional theory (DFT) calculations are used to evaluate the rates of isomerization reactions, which are then implemented in a Monte-Carlo simulation to identify the dominant fragmentation channels (Sect.~\ref{sec:theo}). In Sect.~\ref{sec:results} we detail both the experimental and theoretical results. As explained in Sect.~\ref{sec:mc_assum}, we have restricted our analysis on the isomerization to reactions exclusively involving hydrogen hopping, without modifications on the carbon skeleton. Considering these simplifications, we have limited our results to the first four hydrogen losses. As dehydrogenation progresses further, rearrangements of the carbon structure can affect the validity of our results. We discuss our results in the context of previous work and its astrophysical implications in Sect.~\ref{sec:discussion}. Finally, the main conclusions and possible directions for future research are summarized in Sect.~\ref{sec:conclusions}.

\section{Experimental Methods}
\label{sec:experiment}

This work encompasses nine PAHs, whose chemical formulas and structures are summarized in Fig.~\ref{fig:pahs}. These PAHs were selected as they represent a diverse range of sizes and edge structures and are commercially available. We have focused on medium to large sized PAHs, as they are thought to be more representative of interstellar PAHs. A number of small PAHs have also been included within our sample, in order to compare and validate our results against previous work.

We performed the current experiments with our instrument for photodynamics of PAHs (i-PoP) in the Sackler Laboratory for Astrophysics at Leiden Observatory. A more in-depth description of the system has been provided in \citet{zhe14a}. Relevant details, together with the specific conditions of the current set of experiments, are provided here. The set-up consists of a quadrupole ion trap (QIT) connected to a time-of-flight mass spectrometer (TOF-MS), in order to confine and irradiate precursors and to analyze their photoproducts. Commercially obtained samples of each molecule studied (Fig.~\ref{fig:pahs}) were introduced into an oven one at a time. The oven was then placed under vacuum in the same chamber as the QIT, and subsequently heated up to the sublimation temperature of the selected precursor species. Typical values range from 315~K (triphenylene) to 600~K (dicoronylene). The working pressure of the QIT chamber is $\sim$$5.0\times 10^{-7}$~mbar.

\begin{figure}
  \centering
  \includegraphics[width=\columnwidth]{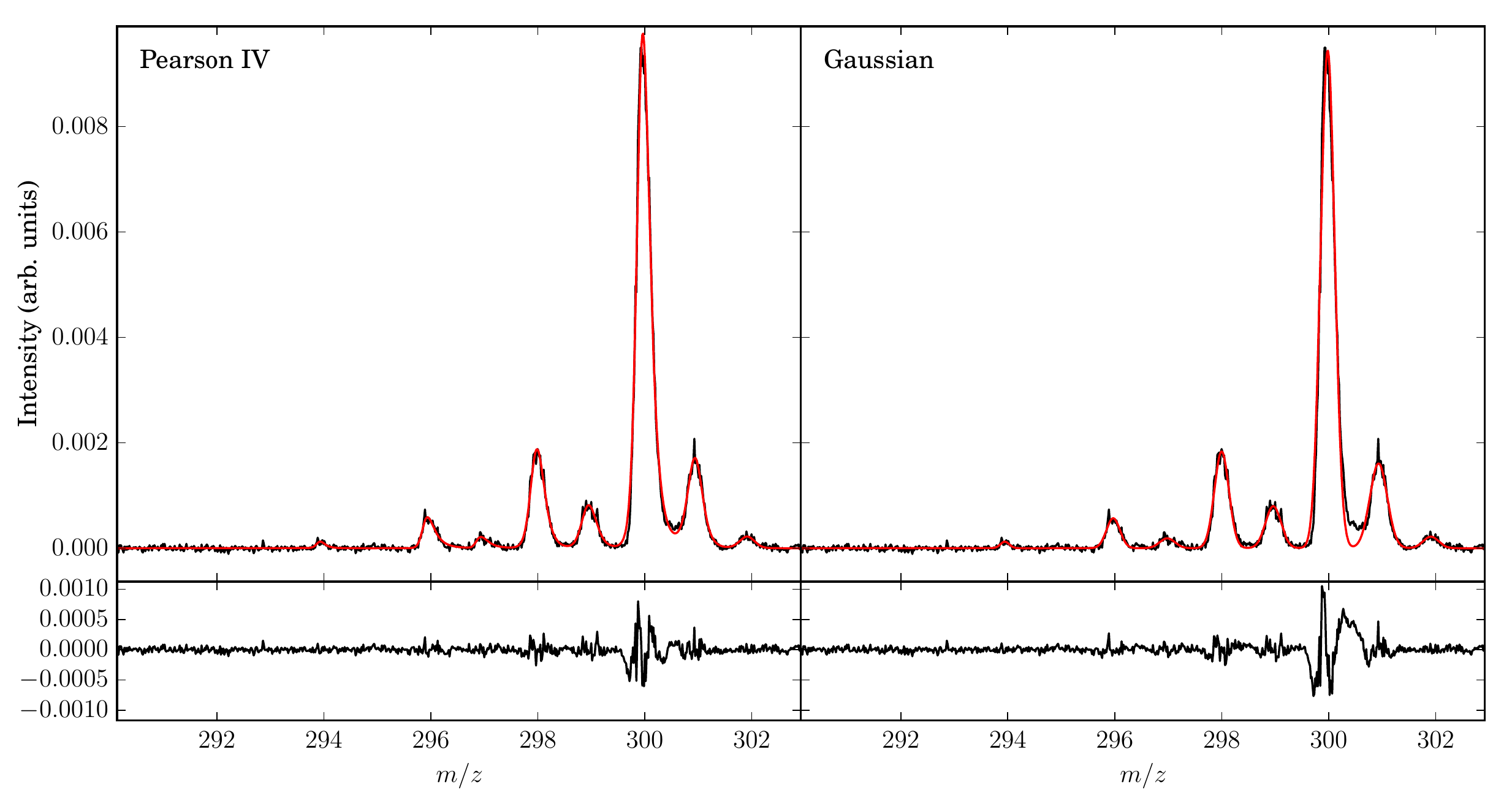}
  \caption{Comparison of a Pearson IV fit (left) to a Gaussian fit (right) on the same coronene mass spectrum. The lower panel shows the difference between the fit and the experimental result.}
  \label{fig:comp}
\end{figure}

Each operation cycle consisted of ionization, QIT filling, mass isolation and irradiation. Ionization of the gas-phase molecules was achieved via electron impact, which also induced partial dehydrogenation. The ions thus produced were then guided into the QIT using ion-optics plates. After filling the QIT for 3~s, the ion-optics voltages were set so as to prevent additional molecules from reaching the trap. Helium gas was injected continuously into the QIT chamber to cool down the ions and thus reduce the size of the ion cloud, increasing the overlap between it and the laser beam. A $\sim$650~$\mu$s long SWIFT pulse was then applied to the end caps of the QIT in order to isolate the isotopically pure parent species. The efficiency of the SWIFT mass-selection varied for different molecules, and it generally worsened as molecular mass increased. The selected ions were then irradiated with a Nd:YAG pumped dye-laser, with a repetition rate of 10~Hz. We used DCM as a dye, tuning the dye-laser to produce radiation at 656~nm ($\sim 2$~eV). By using photons of such wavelength the internal energy of the precursor molecule can build up slowly, thus increasing our sensitivity to the lower energy fragmentation channels. The irradiation time was controlled by a shutter outside the main vacuum chamber and set to 0.3~s, thus allowing three laser pulses. Finally, an extraction pulse sent the photoproducts into the TOF-MS for detection. The final mass spectra were then created by averaging seventy of the previously described cycles at a given laser fluence.

In the case of the largest PAHs studied here (HBC and dicoronylene) the SWIFT pulse was not selective enough to eliminate the isotopic peaks, meaning that in these cases we aimed at reducing the electron-gun induced fragmentation rather than full removal of isotopic peaks. For smaller PAHs, the mass isolation was still not perfect and small levels of fragment and isotopic peaks remained, but did not influence our analysis. Independently of the SWIFT pulse efficiency for each species, the leftover isotopic contribution was arithmetically removed by calculating the ratio of the first isotopic peak to the parent peak and proportionally removing this contribution from the mass spectra. The lowest mass peak observed for a specific PAH cation does not have \ce{^13C} contributions, so the isotopic ratio is multiplied by the intensity of this peak and the result is removed from the immediately higher mass peak. This was then repeated all the way to the parent and isotopic peak.

The intensity of each mass peak in the spectra was calculated by fitting each individual peak using a Pearson IV function. We chose a Pearson IV function to account for the fact that the mass peaks are not perfectly symmetric, and the additional parameters improve the correspondence between the data and the fit. A comparison of the results of this fit to a Gaussian fit to the same mass spectrum is presented in Fig.~\ref{fig:comp}. The decrease in the residuals is noticeable, particularly in the case of high intensity peaks. We estimated the error related to this integration by calculating the root-mean-square over the residual within one standard deviation from the center.

\section{Theoretical methods}
\label{sec:theo}

\subsection{Density Functional Theory calculations}
\label{sec:dft}

Density Functional Theory (DFT) calculations were performed to interpret the results of the experiments.  Intermediate and transition state structures were investigated with B3LYP/6-31G(d,p) using the quantum chemistry software Gaussian~09 \citep{fri09}. Transition state structures were generally found with the help of the Berny algorithm; for more difficult cases  the Synchronous Transit-Guided Quasi-Newton (STQN) method \citep{STQN1,STQN2} was used. Vibrational analysis was performed to verify the nature of the structure --- i.e.\ no imaginary frequency for intermediate states, and one imaginary frequency for transition states --- and to obtain zero-point vibrational energies (ZPEs). Bond dissociation energies (BDEs) were calculated as the difference in the total energy (electronic energy and ZPE) between products and reactants.

\begin{figure*}[t!]
  \centering
  \includegraphics[width=\textwidth]{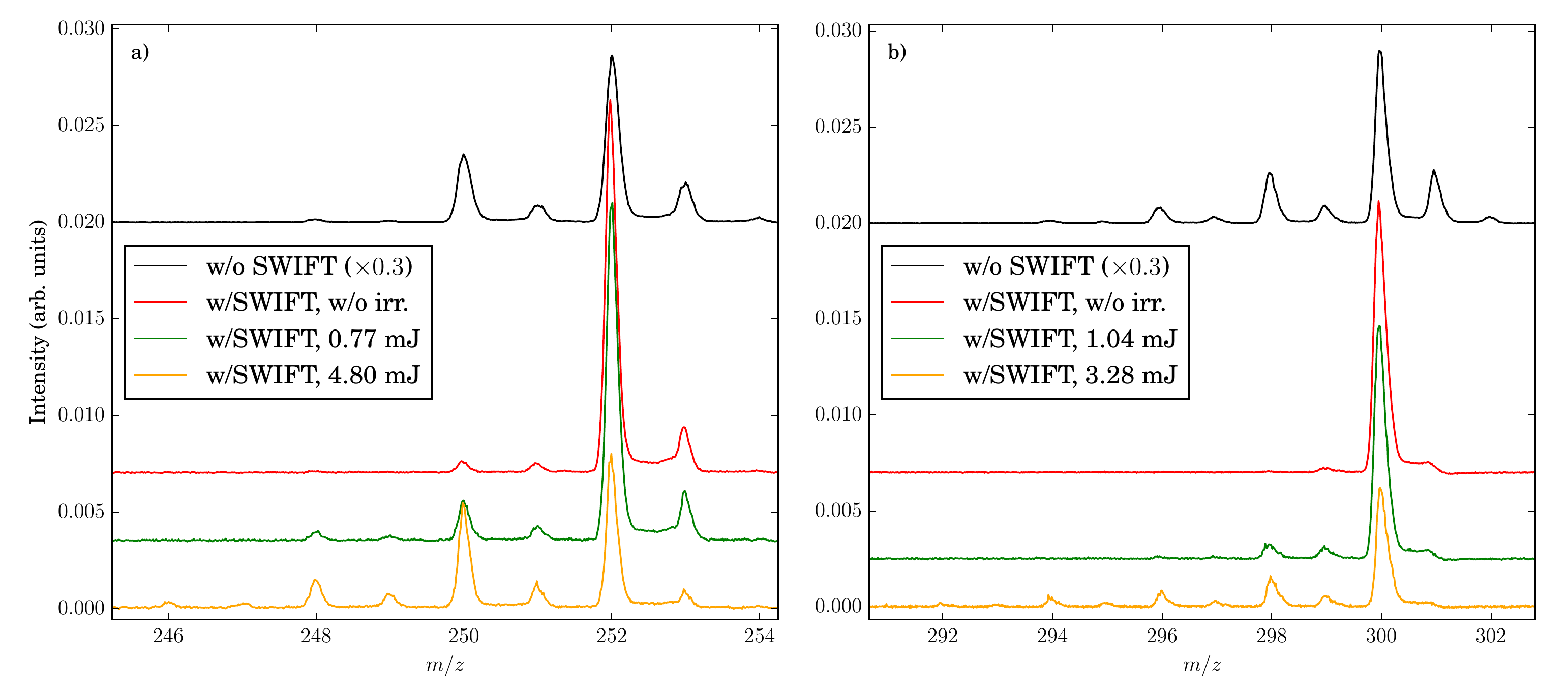}
  \caption{Comparison of the mass spectra of perylene (left) and coronene (right) for different experimental conditions. In black is the spectrum before the SWIFT pulse and without irradiation, which shows electron induced fragmentation (intensity is scaled by 0.3). In red is the SWIFTed and non-irradiated mass spectrum. The green  and yellow traces show the mass spectra, after SWIFT pulse and irradiation, for representative low   and high laser energy, respectively.}
  \label{fig:massspec}
\end{figure*}

To evaluate the accuracy of our calculations, we compared our results to previous experiments or, when these were not available, to high level calculations on smaller and comparable systems. In the case of naphthalene cation (\ce{C10H8+}), barriers for H-hopping reactions calculated at B3LYP/6-31G(d,p) level were found to be overestimated by an average of 0.12~eV  and 0.15~eV when compared with more accurate but highly-expensive methods such as CBS-QB3 \citep{mon00} and G3 \citep{bab00,dya06}. The experimental C--H BDEs for benzene (\ce{C6H6}) and for the two different C--H bonds in naphthalene (\ce{C10H8}) are 4.76 eV \citep{bau98} and 4.87/4.89 eV \citep{ree00}; the values calculated with B3LYP/6-31G(d,p) are 4.80 and 4.80/4.81 eV, with relative errors inferior to 2\%. This comparison validates the use of DFT for our systems.

\subsection{Reaction Rates}
\label{sec:rates}

For each reaction studied with DFT the rate was calculated using Rice-Ramsperger-Kassel-Marcus (RRKM) theory \citep{baer-hase} in the form
\begin{equation}
k(E) = \frac{W^*(E-E_0)}{h\rho(E)},
\end{equation}
where $h$ is the Planck constant,  $W^*$ is the sum of states of the transition state of the reaction, $\rho$ is the density of states of the reactant and $E_0$ is the energy barrier of the reaction, calculated with DFT.  Both the sum of states and density of states were calculated from the DFT normal modes using the \texttt{densum} program from the MULTIWELL suite \citep{Multiwell1,Multiwell2}.
Enthalpy of activation at $T=1000$~K ($\Delta S_{1000}$) was also calculated for each reaction using
\begin{equation}
    \Delta S = k\ln\left[\frac{\Pi\Phi_i^*}{\Pi\Phi_i}\right]+\left(\frac{E^*-E}{T}\right),
\end{equation}
where $E$ and $\Phi$ are the energy and the molecular vibrational partition function of the parent molecule, respectively, while $E^*$ and $\Phi^*$ are the same properties, but for the transition state.

\subsection{Monte-Carlo Simulation}
\label{sec:montecarlo}

In order to model the dehydrogenation process, we have developed a Monte-Carlo code that considers the different hydrogen hopping and fragmentation channels. The rates for these reactions are molecule dependent and calculated using the methods described in the two previous sections. The possible reactions (aromatic H-loss, aromatic \ce{H2}-loss, formation of an aliphatic group, hydrogen hopping to a tertiary\footnote{In the text of this paper we use the adjective tertiary to describe the edge carbon atom bridging two rings.} carbon or emission of an IR photon, in the case of the initial structure) are weighted according to their degeneracies, e.g.\ in the base case of coronene, aromatic H-loss can happen from twelve different positions.The molecular structure of the unperturbed PAH (as in Fig.~\ref{fig:pahs}) is given as the initial configuration when running the Monte-Carlo simulations.  The time-step is adjusted, taking a tenth of the reciprocal of the sum of all the rates, so that there is 90\% chance of nothing happening during each time-step. The weighted rates are converted into a probability by using,
\begin{equation}
p_i = 1-\exp(-k_i\Delta t),
\end{equation}
where $p_i$ and $k_i$ are, respectively, the probability and the rate of the $i$-th reaction and $\Delta t$ is the time-step. The outcome of the model is randomly determined while the structure, its internal energy and the time are updated after each event.

\begin{figure*}[ht!]
  \centering
  \includegraphics[width=\textwidth]{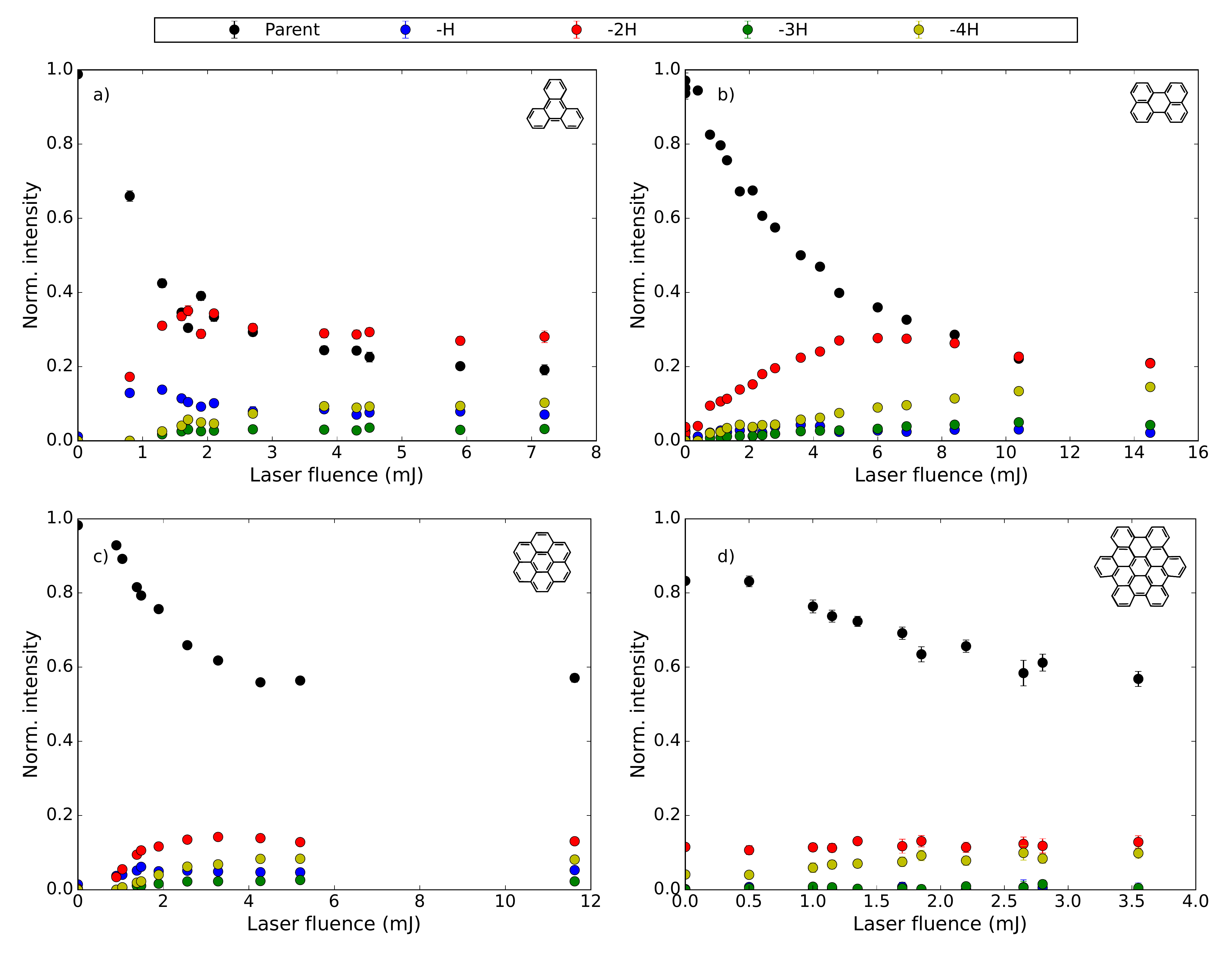}
  \caption{Normalized intensity for the first 4H losses in PAHs with a single type of peripheral hydrogens. (a) Triphenylene (\ce{C18H12+}), with only quarto hydrogens connected by bay-regions; (b) Perylene (\ce{C20H12+}), with only trio hydrogens, also connected by bay-regions; (c) Coronene (\ce{C24H12+}), with only duo hydrogens connected through tertiary carbons; (d) Hexabenzocoronene (\ce{C42H18+}), containing only trio hydrogens connected by bay-regions. In black is shown the intensity of the corresponding parent molecule, while in blue, red, green and yellow are displayed the fragments corresponding to one, two, three and four hydrogen losses respectively.}
  \label{fig:sedge}
\end{figure*}

The code can be run in two different settings. In one, the initial internal energy of the molecule is set and left to evolve for a fixed amount of time in order to determine the AE of the fragments and the main dissociation channels involved. In the second setting, the code simulates the experimental conditions in our set-up. In this variation, an additional photon absorption rate is included, based on the cross-sections derived by \citet{mal07} and the laser fluence. Photon absorption is limited to the first 2~ns, as per the pulse duration. Given the 10~Hz repetition rate of the laser, the molecule is left to evolve for 0.1~s before the next photon absorption phase. In this manner, the three pulses are simulated and the molecule is permitted to evolve for one additional second in order to account for the time between the last pulse and the extraction into the TOF. In this last setting, the partial overlap of the ion-cloud with the laser beam has been included by adding the possibility (ranging from 20 to 40\%) that in a given pulse the molecule will not be able to absorb photons, as it stays outside the beam area. In both formats, the simulation is run for 2500 independent trials per data point (internal energy or laser fluence, respectively) in order to have proper statistics.

\section{Results}
\label{sec:results}

In the next subsections we describe the results of our work. Detailed experimental results for each molecule are summarized in Sect.\ref{sec:exp_res}. The results of DFT calculations, in particular the barriers for different channels for H- and \ce{H2}-loss are presented for a subgroup of the studied molecules in Sect.~\ref{sec:theo_res}. Finally, in Sect.~\ref{sec:mc} we present the results of the Monte-Carlo simulations applied to three molecules of our sample --- perylene, coronene and ovalene --- discussing their limitations and applicability to the experimental results of our sample.

\subsection{Experiments}
\label{sec:exp_res}

Figure~\ref{fig:massspec} shows examples of mass spectra for perylene and coronene. A general property found in the fragmentation pattern of PAHs studied here is that even mass peaks are enhanced with respect to the preceding odd peak, as noted in previous work \citep{eke98,zhe14a,job14}. Figure~\ref{fig:massspec} shows also how the odd-to-even ratio in perylene (left panel) increases both as the dehydrogenation moves along and as the laser fluence is increased. For coronene (right panel), this ratio appears to be constant, independently of the hydrogenation state. A general result of our study is that the variation of the odd-to-even ratio with dehydrogenation (1H/2H, 3H/4H etc.) is related to the edge structure of the PAH molecule. We also find that the size of the molecule plays an important role in determining the odd-to-even ratio; large ($N_\mathrm{C} \geq 32$) molecules show a significant decrease in the odd to even hydrogen ratio at all hydrogenation states. Eventually, the size effect ends up washing away the differences arising from the edge structure.

\begin{figure}[t!]
  \centering
  \includegraphics[width=\columnwidth]{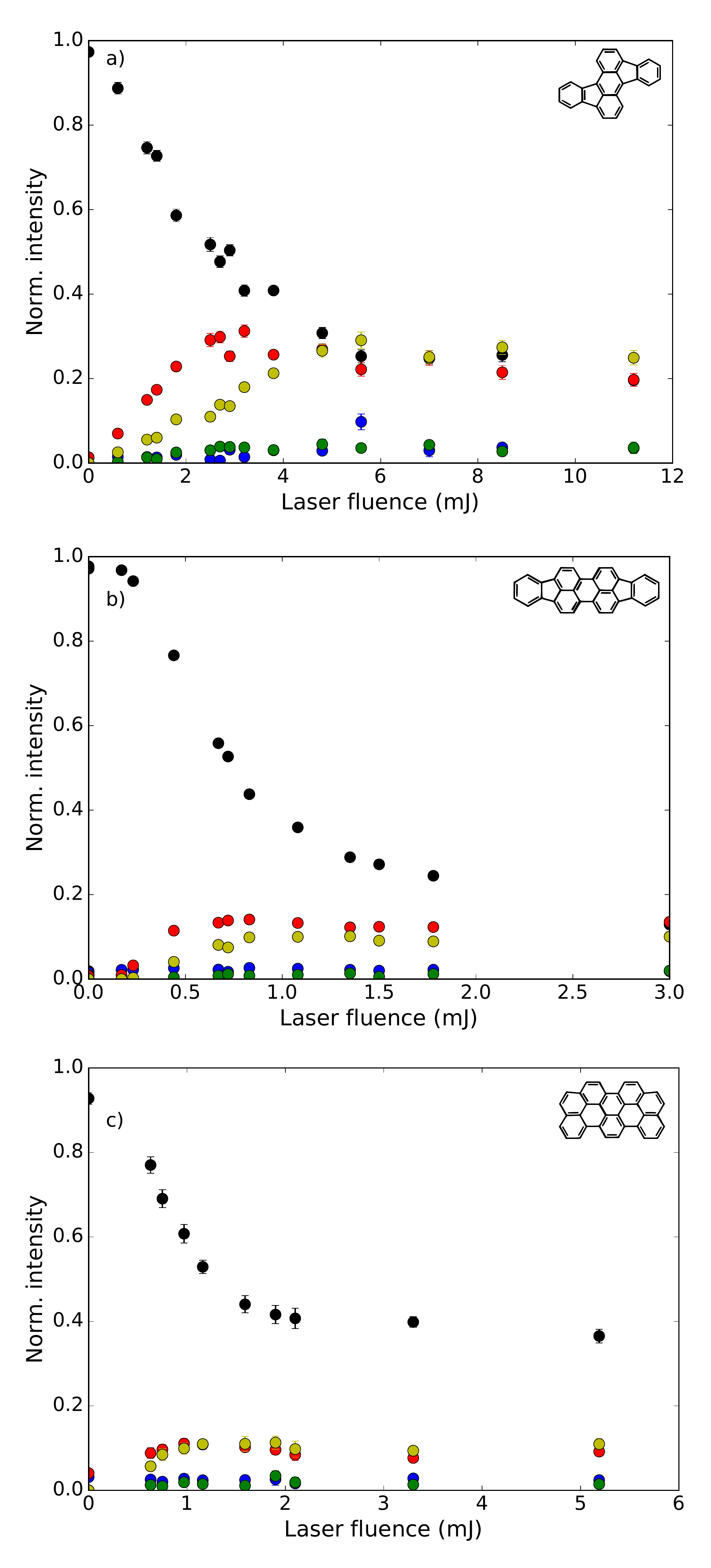}
  \caption{Normalized intensity for the first 4H losses in PAHs with multiple types of peripheral hydrogens. The colors follow the same scheme as Fig.~\ref{fig:sedge}. (a) Rubicene (\ce{C26H14+}), with trio and quarto hydrogens connected by bay-regions; (b) Periflanthene (\ce{C32H16+}), with trio and quarto hydrogens, also connected by bay-regions; (c) TBP (\ce{C34H16+}), with duo and quarto hydrogens connected through tertiary carbons and bay regions.}
  \label{fig:medge}
\end{figure}

In all our experiments the level of dehydrogenation seems to converge at high laser fluence, with the particular value depending on the molecule. We ascribe such behavior to a partial overlap between the laser beam and the ion cloud, which prevents part of the ions  to dissociate. This is further supported by the fact that not only the parent molecule fragmentation stabilizes, but also the fragments themselves do not vary in intensity after that point.

\begin{figure}[t!]
  \centering
  \includegraphics[width=\columnwidth]{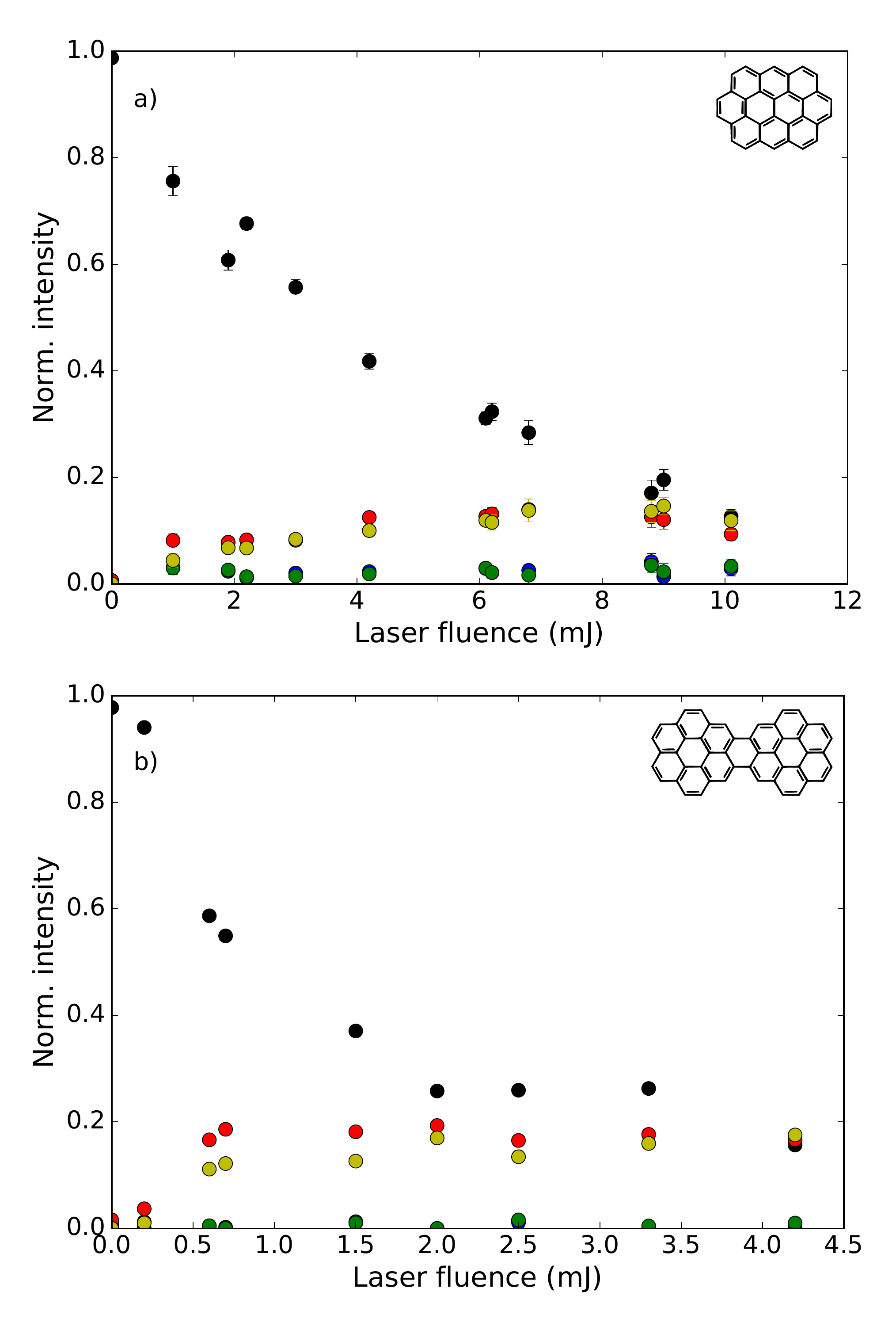}
  \caption{Normalized intensity for the first 4H losses as function of the laser fluence in PAHs with duo and solo hydrogens. The colors follow the same scheme as Fig.~\ref{fig:sedge}. (a) ovalene (\ce{C32H14+}); (b) Dicoronylene (\ce{C48H22+}), which has two bay regions.}
  \label{fig:solo}
\end{figure}

Figure~\ref{fig:sedge} shows breakdown diagrams as function of laser fluence for molecules containing a single type of edge hydrogen. Edge structure appears to drive the changes observed in the different odd and even peak intensities in the case of small molecules ($N_\mathrm{C} \leq 32$). Triphenylene is the smallest molecule considered in our sample (TPH; \ce{C18H12+}) and contains only quarto hydrogens. Only losses up to four hydrogen atoms are detected as observed  by \citet{eke98}. Figure~\ref{fig:sedge}a shows that, while at low laser fluence H-loss and 2H-loss have a similar intensity, the 2H-loss mass peak clearly dominates at 2~mJ. A similar pattern is observed for 3H-loss and 4H-loss, although the laser fluence at which 4H-loss dominates is between 3 and 4~mJ, where H-loss and 4H-loss have almost the same intensity.

\begin{figure*}[t!]
  \centering
  \includegraphics[width=\textwidth]{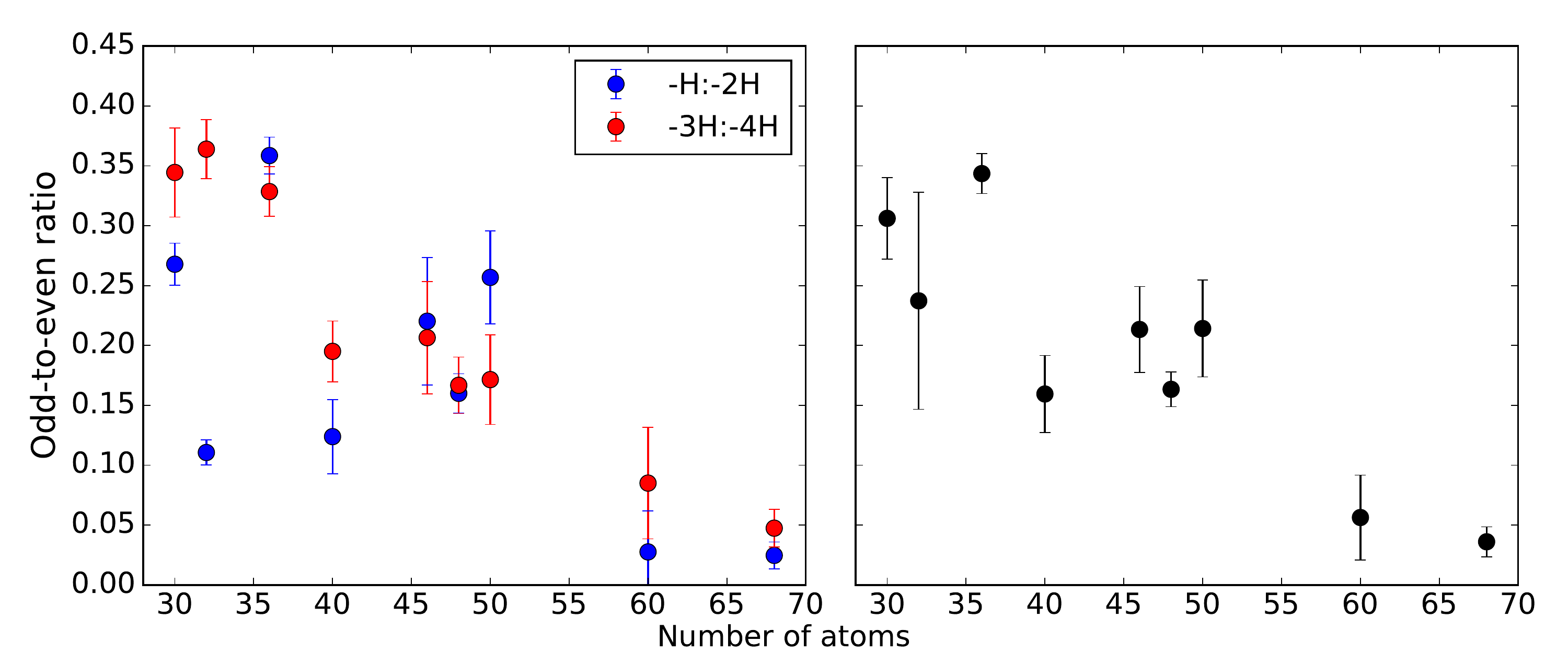}
  \caption{Average odd-to-even ratio (up to 4H losses are considered) for all PAHs studied here as a function of their size in terms of the total number of atoms. The left panel details the variations for progressive dehydrogenation steps (-H/-2H and -3H/-4H) within each PAH. The right panel shows the average of both odd-to-even ratios --- i.e., (-H/-2H$+$-H/-2H)$/2$.}
  \label{fig:size}
\end{figure*}

Perylene (Pery; \ce{C20H12+}, Fig.~\ref{fig:sedge}b) has only trio hydrogens and up to six hydrogen losses are detected in our experiments, two less than what was observed by \citet{eke98}. The odd to even ratio in successive H-losses of perylene increases as the dehydrogenation progresses. The 2H-loss mass peak is much more prominent than that for H-loss. Additionally, both the H-loss and 3H-loss intensities are nearly the same starting at 4~mJ, while for triphenylene both intensities remain different along the whole laser fluence range. Furthermore the 4H-loss channel overcomes the H-loss channel early on, and the ratio of -3H/-4H fragments is higher than the -H/-2H fragment ratio.

In coronene (Coro; \ce{C24H12+}, Fig.~\ref{fig:sedge}c) the fragmentation pattern resembles that of triphenylene. Coronene has only duo hydrogens and a maximum of eight hydrogen atoms are lost within the range of laser fluence explored here. \citet{eke98} noted that coronene reaches full dehydrogenation under their experimental conditions. In our experiments we find that, at laser fluence $\leq$~1~mJ, the H-loss and 2H-loss channels have similar intensities. As the laser fluence increases, the 2H-loss mass peak quickly dominates over the H-loss. Additionally, H- and 3H-loss peak intensities are present at all fluences. Unlike in triphenylene, 4H-loss eventually surpasses the H-loss channel at around 3~mJ. Odd-to-even ratios -H/-2H and -3H/-4H remain similar to each other.

HBC (\ce{C42H18+}; Fig.~\ref{fig:sedge}d), as perylene, contains only trio hydrogens and it is significantly larger than the molecules considered up to here. The dehydrogenation of HBC has  been reported by \citet{zhe14a}; they found the same qualitative behavior observed here for perylene -- the further the dehydrogenation progresses, the larger the increase on the successive odd-to-even ratio. However, such behavior is less pronounced than what we find in perylene and becomes noticeable as HBC approaches complete dehydrogenation. One clear difference for HBC with respect to the previous molecules is that the odd hydrogen loss channels are almost non-existent. While the 2H-loss intensity appears constant, it must be noted that in this case the SWIFT isolation was the least efficient, with a large amount of electron impact fragments remaining from the beginning. 

PAHs with multiple edge structures (Fig.~\ref{fig:medge}) show a mixed behavior, corresponding to that of the edge structures present. The main factor driving the odd-to-even ratio is still related to the molecular size. Rubicene (Rub; \ce{C26H14+}, Fig.~\ref{fig:medge}a) is the smallest molecule in our sample with a mixture of peripheral hydrogen types, in this case trios and quartos. Additionally, it has pentagonal rings along with the more typical hexagons. We detect losses of up to 8 hydrogen atoms, although the decrease of intensity of the parent peak is much more pronounced than that observed for coronene. As was the case for perylene, the H- and 3H-loss intensities are nearly constant for all laser fluences, although with a lower intensity. At laser fluences below 3~mJ the odd-to-even ratios for -H/-2H and -3H/-4H behave similarly to those observed for perylene. This situation changes at the high end of the fluence range, with the fragmentation pattern showing a closer resemblance to that of coronene, with both ratios having nearly the same values.

Periflanthene (PF; \ce{C32H16+}, Fig.~\ref{fig:medge}b) has a combination of duo and quarto hydrogens and, as rubicene, contains two pentagonal rings in its structure. Up to twelve hydrogen losses are observed. The intensity behavior with laser fluence qualitatively resembles that of coronene, although odd hydrogen losses are lower in intensity. Also the 4H-loss channel overcomes the H-loss channel at a much lower laser fluence.

TBP (\ce{C32H16+}; Fig.~\ref{fig:medge}c) has mostly duo hydrogens, but there are two trio structures as well, with a mixture of bay-regions and tertiary carbons as the connectors. Overall, the dissociation pattern is close to that of periflanthene, displaying the same low intensity for the odd hydrogen losses. The most noticeable difference stems from the fact that 2H- and 4H-loss channel intensities are nearly identical for most of the laser fluence range, a characteristic observed at high laser fluence in rubicene, HBC and, to a lesser extent, in perylene.

\begin{figure*}[ht!]
  \includegraphics[width=\textwidth]{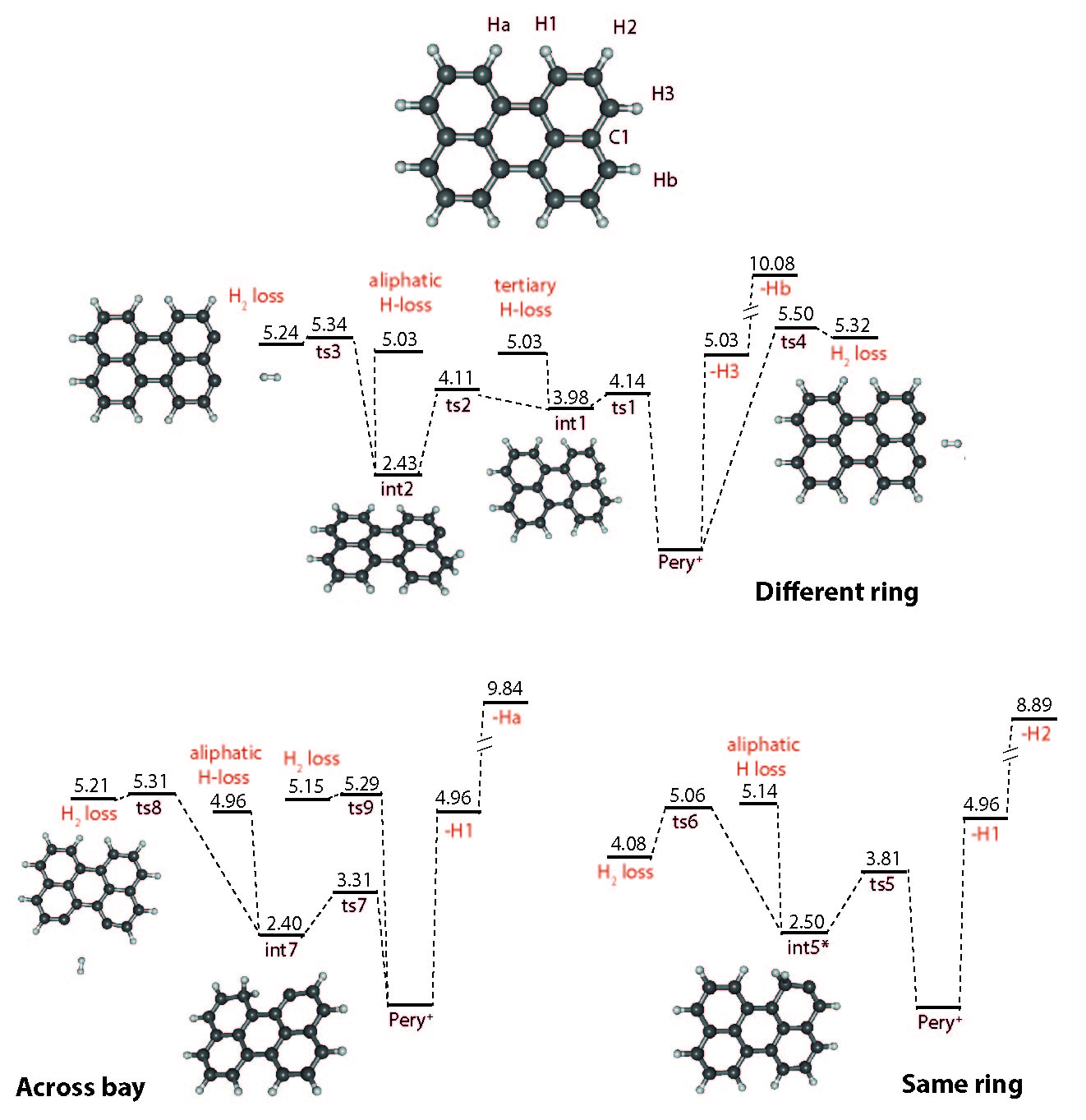}
  \caption{Energy scheme for the perylene cation (top-left), showcasing the different H-hop and H-loss channels available. Isomerization/fragmentation pathways are classified according to whether they involve adjacent rings separated by a single tertiary carbon (top-right), neighboring rings separated by bay-regions (bottom-left) or a single ring (bottom-right). Intermediate and transitions states are labeled as int and ts, respectively. Loss channels are highlighted in red and all energies are given in eV with respect to that of the perylene cation.  The * highlights an intermediate state  that can have an energy of 3.20~eV if the aliphatic group is formed by hopping of H1 or H3 towards H2, as opposed to H2 hopping towards H1 (illustrated here) or H3 (see main text).} 
  \label{fig:energy_scheme}
\end{figure*} 

Ovalene (Ova; \ce{C32H14+}, Fig.~\ref{fig:solo}a) has a similar behavior as TBP (a molecule with the same size but a less compact structure) although the fragmentation of the parent progresses much further. Ovalene has mainly duo hydrogens but also two solo hydrogens and looses up to twelve hydrogen atoms. The odd hydrogen loss channels are, as for TBP, lower than what has been observed for other molecules, resulting in constant odd-to-even ratios as dehydrogenation progresses. The largest molecule in our sample, dicoronylene (DC; \ce{C48H22+}, Fig.~\ref{fig:solo}b), is composed of mostly duo hydrogens with four solos connected by bay regions. This makes it similar to ovalene in terms of the proportion of edge structures. However, the odd hydrogen fragments are effectively zero, as in HBC, the only other molecule in our sample with more than 40 C-atoms.

In summary, variations in the odd-to-even ratio as the hydrogenation level of the molecule decreases appear to correlate with the number of trios for small molecules (Fig.~\ref{fig:size}, left panel), with the largest difference by far observed in the fragmentation pattern of perylene. Molecules with no trios, on the other hand, show a nearly equal -H/-2H and -3H/-4H ratios. However, as the size increases, the intensity of peaks corresponding to odd hydrogen fragment becomes progressively lower and the difference in successive odd-to-even ratios becomes less pronounced, as can be seen from the average ratios (Fig.~\ref{fig:size}, right panel) This is particularly evident as the molecules approach 50 C-atoms, for which the odd H products are barely detected. The effect of trios and the decrease in the intensity of odd peaks with increased molecular size are explored next using a combination of DFT calculations and Monte-Carlo simulations.

\subsection{Dehydrogenation channels and isomerization}
\label{sec:theo_res}

We investigated the possible dehydrogenation channels for the PAH cations studied experimentally by calculating a) bond dissociation energies (BDE) for H cleavages at different sites and b) transition  and intermediate states involved in \ce{H2} formation as function of dehydrogenation. Calculation shows that BDEs for H cleavage are independent of both molecular size and degree of dehydrogenation of the molecule. The edge structure of the molecule, namely if the edge H atoms are alone (solo) or arranged in two, three or four (duo, trio and quarto, respectively, Fig.\ref{fig:pahs}) affects the dissociation energy. When an edge hydrogen is alone (solo) an average of $\sim$4.8~eV are needed to remove it. In a duo group, the removal of the first H-atom (both are equivalent) is comparable to that of a solo H, $\sim$4.8~eV. The same energy is required to remove the first hydrogen in trios and quartos if this H occupies an external position in the ring. In the case of the central hydrogens (one for the trio and two for the quarto), the energy required for the bond cleavage is somewhat higher, at $\sim$5.0~eV. These results are in agreement with previous theoretical studies \citep{aihara96}.

The BDE for the second hydrogen atom in the same ring, independently on whether it belongs to a duo, trio or quarto, needs $\sim$3.8~eV when it is next to the first dehydrogenated site. The lower energy with respect to the first hydrogen removal is ascribed to the spin pairing of the adjacent carbons in the nearby site, which creates a triple bond as the lowest energy structure, with a doublet electronic state \citep{job14}. In trios and quartos, when the second hydrogen is removed from a position not immediately adjacent to the dehydrogenated carbon, the BDE is of the order of 4.0~eV. In this case a $\sigma$ bond forms between the two dehydrogenated sites, creating a cyclopropenyl unit which stabilizes cationic molecules \citep{tri17a}. Removal of the third hydrogen in a trio leaves an unpaired electron and thus requires $\sim$4.9~eV. The same energy is required in a quarto if the third  H loss is non-adjacent to a dehydrogenated site. In the opposite case, it will require again only $\sim$3.8~eV. The final hydrogen removal in a quarto is nearly independent of the position and it requires 4.2~eV.

C--H BDEs are determined mostly by the local alteration of the molecular structure near the dehydrogenation site. This results in the invariance of the BDEs as dehydrogenation progresses and in the fact that when a group, e.g.\ a trio, is half dehydrogenated, removal of H from another group will require 4.8~eV. Small fluctuations of the energies are observed in the presence of bay-region. For instance, the BDE of the first C--H bond will require slightly lower energy (0.2~eV) when it is removed within the bay region rather than at other positions. This is due to the release of steric hindrance in the bay region \citep{cas18}.

It is interesting to notice that the different energies for sequential H-loss in the case of trio hydrogens could explain the differences observed in the experiments for small molecules like perylene, but not why such differences apparently disappear as molecular size grows. This observation pushed us to explore other pathways leading to hydrogen loss. Given that calculations for the molecules in the sample show that the barriers for these reaction paths are independent of molecular size and dehydrogenation level, we chose perylene to describe these channels and their representative energy values (Fig.~\ref{fig:energy_scheme}). Indeed, perylene contains both bay-region and single tertiary carbon connections between rings so all the pathways investigated are present at once.

Direct molecular hydrogen formation (not involving aliphatic C--H bonds) can proceed through the stretching of the C--H bonds in rings separated by a tertiary carbon (C1, Fig.~\ref{fig:energy_scheme}, top-right) as found for zig-zag edges PAHs \citep{job03,Paris14,Chen15}. \ce{H2} formation in this case involves the hydrogens labeled H3 and Hb, with a transition state (ts4) of 5.50~eV. In rings separated by bay-regions (Fig.~\ref{fig:energy_scheme}, bottom-left), C--H stretch of the nearby hydrogen bonds (involving H1 and Ha) can also lead to \ce{H2} formation, in this case with a lower transition state of 5.29~eV (ts9).

\begin{table}[!t]
    \begin{adjustbox}{max width=\columnwidth}
    \begin{threeparttable}
        \caption{Rates included in Monte-Carlo simulations and parameters for Perylene$^+$.}
        \label{tab:trans_per}
        \begin{tabular}{lddd}
            \toprule
            Transition type & \multicolumn{1}{c}{$E_0$} & \multicolumn{1}{c}{$\Delta E$} & \multicolumn{1}{c}{$\Delta S_{1000}$} \\
            & \multicolumn{1}{c}{(eV)} & \multicolumn{1}{c}{(eV)} & \multicolumn{1}{c}{(J~K$^{-1}$~mol$^{-1}$)} \\ 
            \midrule
            \multicolumn{4}{c}{Hydrogen losses} \\
            Arom. H-loss & 4.96 & 4.96 & 3.40 \\
            Seq. arom. H-loss & 3.93 & 3.93 & 34.35 \\
            Arom. \ce{H2}-loss & 5.50 & 5.32 & 35.36 \\
            Aliph. H-loss (del.) & 2.64 & 2.64 & -4.43 \\
            Aliph. H-loss (loc.) & 1.93 & 1.93 & 20.48 \\
            Aliph. \ce{H2}-loss (del.) & 2.56 & 1.58 & 27.03 \\
            Aliph. \ce{H2}-loss (loc.) & 1.84 & 0.86 & -2.61 \\
            Tert. H-loss & 1.05 & 1.05 & 0.87 \\
            Bay \ce{H2}-loss & 5.29 & 5.15 & 43.74 \\
            \multicolumn{4}{c}{H-roaming inside single ring} \\
            Aliph. form. (del.) & 3.81 & 2.50 & 0.08 \\
            Aliph. form. (loc.) & 3.81 & 3.21 & 0.46 \\
            Aliph. to empty (del.) & 1.27 & -2.50 & -8.09 \\
            Aliph. to empty (loc.) & 0.60 & -3.21 & -13.14 \\
            Aliph. exchange* & 1.50 & 0.00 & 21.14 \\
            Arom. exchange* & 2.39 & 0.00 & -3.71 \\
            \multicolumn{4}{c}{H-roaming to/from tertiary carbon} \\
            Aliph. to tert.* & 1.68 & 1.55 & -8.82 \\
            Tert. to aliph.* & 0.12 & -1.55 & -12.11 \\
            Arom. to tert.* & 4.14 & 3.98 & 5.40 \\
            Tert. to arom.* & 0.16 & -3.98 & -10.62 \\
            \multicolumn{4}{c}{H-roaming across bay-region} \\
            Aliph. form. & 3.31 & 2.40 & -11.77 \\
            Aliph. to empty & 0.91 & -2.40 & -21.80 \\
            Aliph. exchange & 1.82 & 0.00 & -18.86 \\
            Arom. exchange & 1.20 & 0.00 & -11.56 \\
            \bottomrule
        \end{tabular}
        \begin{tablenotes}
            \item *: $E_0$ and $\Delta S$ depend on hydrogenation level of ring involved.
        \end{tablenotes}
  \end{threeparttable}
  \end{adjustbox}
\end{table}

Molecular hydrogen formation and single H-loss are also possible once aliphatic C--H bonds are available (in the form of \ce{CH2} groups) via H-hopping. Within the same ring (Fig.~\ref{fig:energy_scheme}, bottom-right), we have such a situation when a hydrogen atom hops to a nearby occupied site (ts5), creating a \ce{CH2} group adjacent to an empty site (int5). For trios and quartos, the presence of the \ce{CH2} group in the central carbon(s) (that is, H1 or H3 moving into H2) will lead to charge localization. While the transition state involved is the same, the intermediate will be $\sim$0.8~eV higher in energy than for H-hopping in the opposite direction. Once the \ce{CH2} group is formed, a single C--H bond can be severed (requiring an additional 2.64~eV) or \ce{H2} can be formed and released via transition state (ts6) requiring slightly lower energy (2.56~eV with respect to int5). This difference is within the accuracy of our DFT calculations, and for different molecules it can favor H-loss rather than \ce{H2}-loss. 

\ce{H2}-loss can be achieved also in rings separated by tertiary carbons. For perylene (Fig.~\ref{fig:energy_scheme}, top-left), this can happen when H3 hops to the tertiary carbon C1 (int1) through a transition state. From there either H-loss can proceed directly with an additional 1.05~eV or H3 can jump a second time towards Hb (ts2; requiring 0.13~eV). The resulting isomer (int2) now has an aliphatic group adjacent to an occupied carbon, and H-loss requires an additional 2.60~eV. The resulting energies of the transition and intermediate states for this type of H-hopping are in good agreement with previous results \citep{Paris14, Chen15}. The transition state for \ce{H2}-loss (ts3) on the other hand, now needs 2.91~eV above int2 to take place. Finally,  an aliphatic group can form in the bay region if H1 jumps to the neighboring hydrogen Ha, without passing through the tertiary carbons (ts7). This configuration (int5) can again lead to H-loss (2.56~eV above int7) or to \ce{H2}-loss (via ts8, 2.91~eV above int7).

\begin{table}[!t]
  \begin{adjustbox}{max width=\columnwidth}
    \begin{threeparttable}
      \caption{Rates included in Monte-Carlo simulations and parameters for Coronene$^+$.}
      \label{tab:trans_cor}
      \begin{tabular}{lddd}
        \toprule
        Transition type & \multicolumn{1}{c}{$E_0$} & \multicolumn{1}{c}{$\Delta E$} & \multicolumn{1}{c}{$\Delta S_{1000}$}  \\
        & \multicolumn{1}{c}{(eV)} & \multicolumn{1}{c}{(eV)} & \multicolumn{1}{c}{(J~K$^{-1}$~mol$^{-1}$)} \\ 
        \midrule
        \multicolumn{4}{c}{Hydrogen losses} \\
        Arom. H-loss & 4.86 & 4.86 & -6.70 \\
        Seq. arom. H-loss & 3.84 & 3.84 & 46.56 \\
        Arom. \ce{H2}-loss & 5.42 & 5.23 & 24.17 \\
        Aliph. H-loss & 2.49 & 2.49 & 38.80 \\
        Aliph. \ce{H2}-loss & 2.59 & 1.76 & 22.74 \\
        Tert. H-loss & 1.78 & 1.78 & 14.64 \\
        \multicolumn{4}{c}{H-roaming inside single ring}\\
        Aliph. formation & 3.18 & 2.36 & -18.82 \\
        Aliph. to empty & 0.82 & -2.36 & -11.82 \\
        Aliph. exchange & 0.79 & 0.00 & -14.30 \\
        Arom. exchange & 2.71 & 0.00 & 12.07 \\
        \multicolumn{4}{c}{H-roaming to/from tertiary}\\
        Aliph. to tert.* & 1.15 & 0.66 & -12.44 \\
        Tert. to aliph.* & 0.49 & -0.66 & -9.99 \\
        Arom. to tert.* & 3.57 & 3.07 & -17.40 \\
        Tert. to arom.* & 0.50 & -3.07 & -7.83 \\
        \bottomrule
      \end{tabular}
      \begin{tablenotes}
        \item *: $E_0$ and $\Delta S$ depend on hydrogenation level of ring involved.
      \end{tablenotes}
    \end{threeparttable}
  \end{adjustbox}
\end{table}

\begin{table}[!th]
  \begin{adjustbox}{max width=\columnwidth}
    \begin{threeparttable}
      \caption{Rates included in Monte-Carlo simulations and parameters for Ovalene$^+$.}
      \label{tab:trans_ova}
      \begin{tabular}{lddd}
        \toprule
        Transition type & \multicolumn{1}{c}{$E_0$} & \multicolumn{1}{c}{$\Delta E$} & \multicolumn{1}{c}{$\Delta S_{1000}$}  \\
        & \multicolumn{1}{c}{(eV)} & \multicolumn{1}{c}{(eV)} & \multicolumn{1}{c}{(J~K$^{-1}$~mol$^{-1}$)} \\ 
        \midrule
        \multicolumn{4}{c}{Hydrogen losses} \\
        Arom. H-loss (solo) & 4.82 & 4.82 & 15.58 \\
        Arom. H-loss (duo) & 4.84 & 4.84 & 3.93 \\
        Seq. arom. H-loss & 3.83 & 3.83 & 16.71 \\
        Arom. \ce{H2}-loss & 5.26 & 5.15 & 35.43 \\
        Aliph. H-loss (solo) & 2.61 & 2.61 & -2.53 \\
        Aliph. H-loss (duo; loc.) & 1.92 & 1.92 & -8.41 \\
        Aliph. H-loss (duo; del.) & 2.23 & 2.23 & -7.36 \\
        Aliph. H-loss (duo; far) & 2.15 & 2.15 & 13.69 \\
        Aliph. \ce{H2}-loss (solo) & 3.23 & 3.03 & 23.88 \\
        Aliph. \ce{H2}-loss (duo; loc.) & 1.99 & 1.18 & 12.83 \\
        Aliph. \ce{H2}-loss (duo; del.) & 3.01 & 1.45 & 30.02 \\
        Aliph. \ce{H2}-loss (duo; far) & 2.22 & 1.39 & 15.28 \\
        Tert. H-loss & 1.49 & 1.49 & -6.97 \\
        \multicolumn{4}{c}{H-roaming inside single ring}\\
        Aliph. formation (loc.) & 3.60 & 2.92 & 0.69 \\
        Aliph. to empty (loc.) & 0.68 & -2.92 & -11.25 \\
        Aliph. formation (del.) & 3.57 & 2.65 & 1.07 \\
        Aliph. to empty (del.) & 0.92 & -2.65 & -9.82 \\
        Aliph. formation (far) & 3.65 & 2.71 & 0.29 \\
        Aliph. to empty (far) & 0.94 & -2.71 & -9.95 \\
        Aliph. exchange & 0.90 & 0.00 & -13.54 \\
        Arom. exchange & 2.70 & 0.00 & 7.11 \\
        \multicolumn{4}{c}{H-roaming to/from tertiary}\\
        Aliph. (solo) to tert. & 1.65 & 1.48 & -12.57 \\
        Tert. to aliph. (solo) & 0.17 & -1.48 & -11.70 \\
        Arom. (solo) to tert. & 3.98 & 3.75 & -1.40 \\
        Tert. to arom. (solo) & 0.23 & -3.75 & -10.22 \\
        Aliph. (duo) to tert.* & 1.12 & 0.60 & -11.78 \\
        Tert. to aliph. (duo)* & 0.52 & -0.60 & -10.64 \\
        Arom. (duo) to tert.* & 4.07 & 3.36 & 2.53 \\
        Tert. to arom. (duo)* & 0.71 & -3.36 & -5.25 \\
        \bottomrule
      \end{tabular}
      \begin{tablenotes}
      \item *: $E_0$ and $\Delta S$ depend on hydrogenation level of ring involved.
      \end{tablenotes}
    \end{threeparttable}
  \end{adjustbox}
\end{table}

\subsection{Monte-Carlo simulations}
\label{sec:mc}

\subsubsection{Assumptions and Limitation}
\label{sec:mc_assum}

Before moving to the results of the Monte-Carlo simulations, it is useful to describe in detail the rates considered and the assumptions made in modeling the reactions involved in the dehydrogenation process. These assumptions stem from the results of DFT calculations for perylene, coronene and ovalene cations and affect the rates included in the simulations. The rates used for the Monte-Carlo simulations are summarized in Tables~\ref{tab:trans_per} to \ref{tab:trans_ova}. Rates are separated in terms of those involving hydrogen losses, H-roaming within a single ring, jumps to and from tertiary carbons, and hydrogen jumps across bay regions. H-loss can proceed directly from a fully aromatic ring or can have a reduced barrier for rings that have already lost an odd number of hydrogens. Instances where intermediates have a localized charge upon aliphatic formation are indicated in Tables~\ref{tab:trans_per} and \ref{tab:trans_ova}. In the case of ovalene, the formation of the aliphatic in the position immediately next to a solo site has been labeled as delocalized, while the duos which are non-contiguous to the solo have been labeled as ``far" in Table~\ref{tab:trans_ova}. This localization clearly will affect the rates of both the aliphatic H-loss and the \ce{H2}-loss.

\begin{figure*}[p!]
  \centering
  \includegraphics[width=\textwidth]{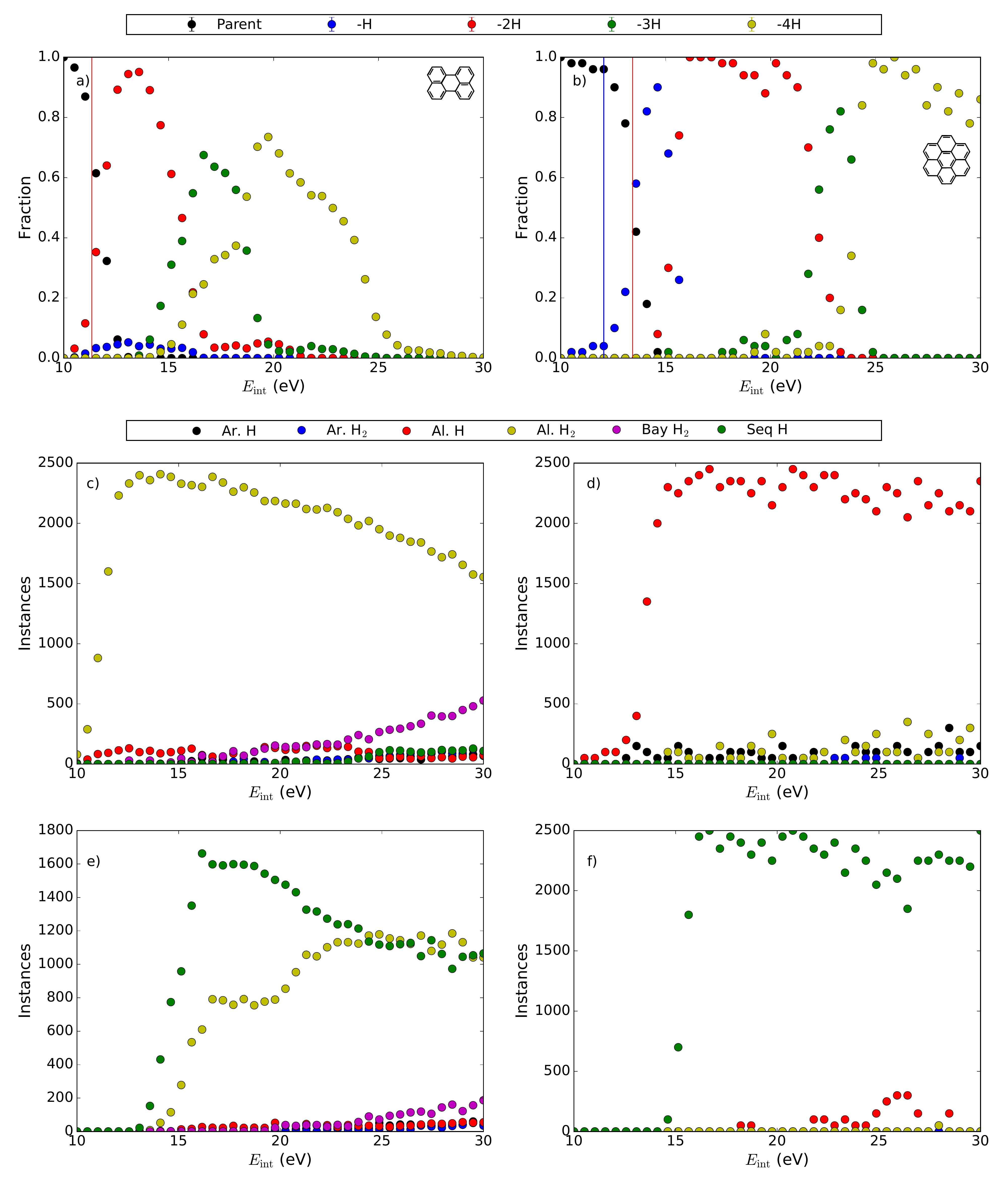}
  \caption{Results of Monte-Carlo simulations of the experiments by \citet{joc94} for the hydrogen loss channels of perylene (a) and coronene (b) after $10^{-4}$~s. The vertical lines indicate the experimentally derived AE for H-loss (blue) and \ce{H2}-loss (red). Panels (c) and (d) show the competition between the different dehydrogenation channels involved in the first fragmentation from perylene and coronene, respectively. Panels (e) and (f) also show the competition between the different channels, but in this case for the second fragmentation step (from perylene after \ce{H2}-loss and from coronene after H-loss).}
  \label{fig:Eint}
\end{figure*}

We considered only hydrogen roaming across the edge of the molecules. H-roaming within a single ring includes reactions that create an aliphatic \ce{CH2} group, with a hydrogen hopping to a nearby occupied site, and its reverse. Aliphatic exchange refers to one of the hydrogens from a \ce{CH2} group moving into a nearby carbon already occupied by a hydrogen atom. The aromatic exchange corresponds to hydrogen atoms jumping to unoccupied neighboring carbon atoms. The exchange reactions within the same ring are affected by the number of empty sites in the ring in question. For instance, in the case of perylene, the aromatic exchange in a trio with two unoccupied sites has $E_0 = 3.30$~eV, while listed in Table~\ref{tab:trans_per} is the value for aromatic exchange when only one site is unoccupied and we have $E_0 = 2.39$~eV.

\begin{figure*}[t!]
  \centering
  \includegraphics[width=\textwidth]{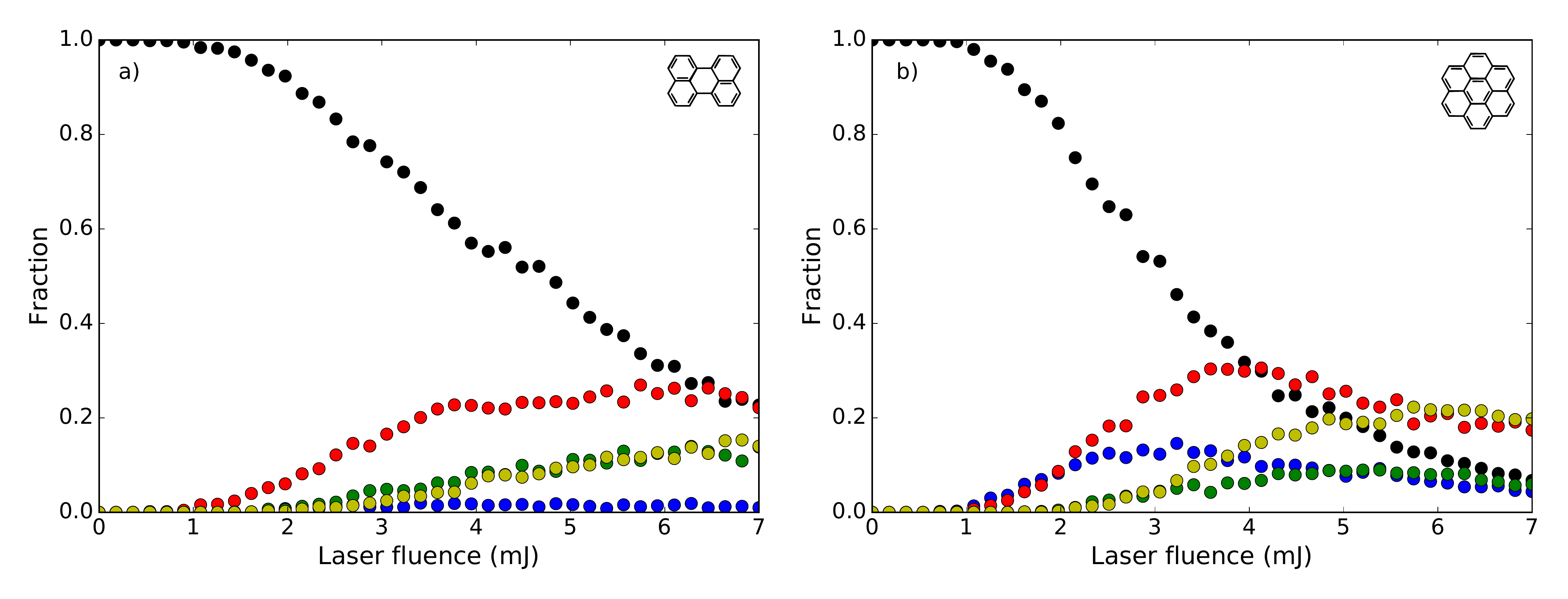}
  \caption{Results of Monte-Carlo simulations of the i-PoP experiments on the fragmentation of perylene (a) and coronene (b) after irradiation with three laser pulses at 656~nm. The colors follow the same scheme as those of the top panels of Fig.~\ref{fig:Eint}.}
  \label{fig:ipop_sim}
\end{figure*}

Hydrogen shifts to or from tertiary carbons also exhibit a change in the barriers depending on the hydrogenation level of the ring involved in the shift. This is especially noticeable for shifts involving aromatic units. Additionally, hydrogen jumps from tertiary carbons are the only reactions found here that are affected by the edge structure in question. For instance, in the case of ovalene, H-hopping into an occupied solo requires $E_0 = 0.17$~eV, while the same reaction into an occupied duo has $E_0 = 0.52$~eV, which has a significant effect in the rate (Table~\ref{tab:trans_ova}).

DFT calculations showed that the barriers for even and odd H-loss and \ce{H2}-loss are insensitive to the degree of dehydrogenation. However, dehydrogenation affects the energy barriers and $\Delta S$ of H-roaming reactions, especially when H is moving along the edge from one ring to the next through a tertiary carbon in the three molecules considered. In all of the other cases the fluctuations are a few percent, thus within the accuracy of the  calculated barriers. Hence we consider all the rates, except the H-roaming across different rings, to be independent of the hydrogen coverage of the molecule.

Additionally, even though H-shifts are known to differ in energy as the distance between the \ce{CH2} and the radical site changes \citep{tri17a}, we decided to consider single rates for hydrogen shifts of similar nature. The energy differences found by \citet{tri17a} in the case of neutral coronene are up to 0.4~eV, which should not significantly affect these rates. 

\begin{figure*}[t!]
  \centering
  \includegraphics[width=\textwidth]{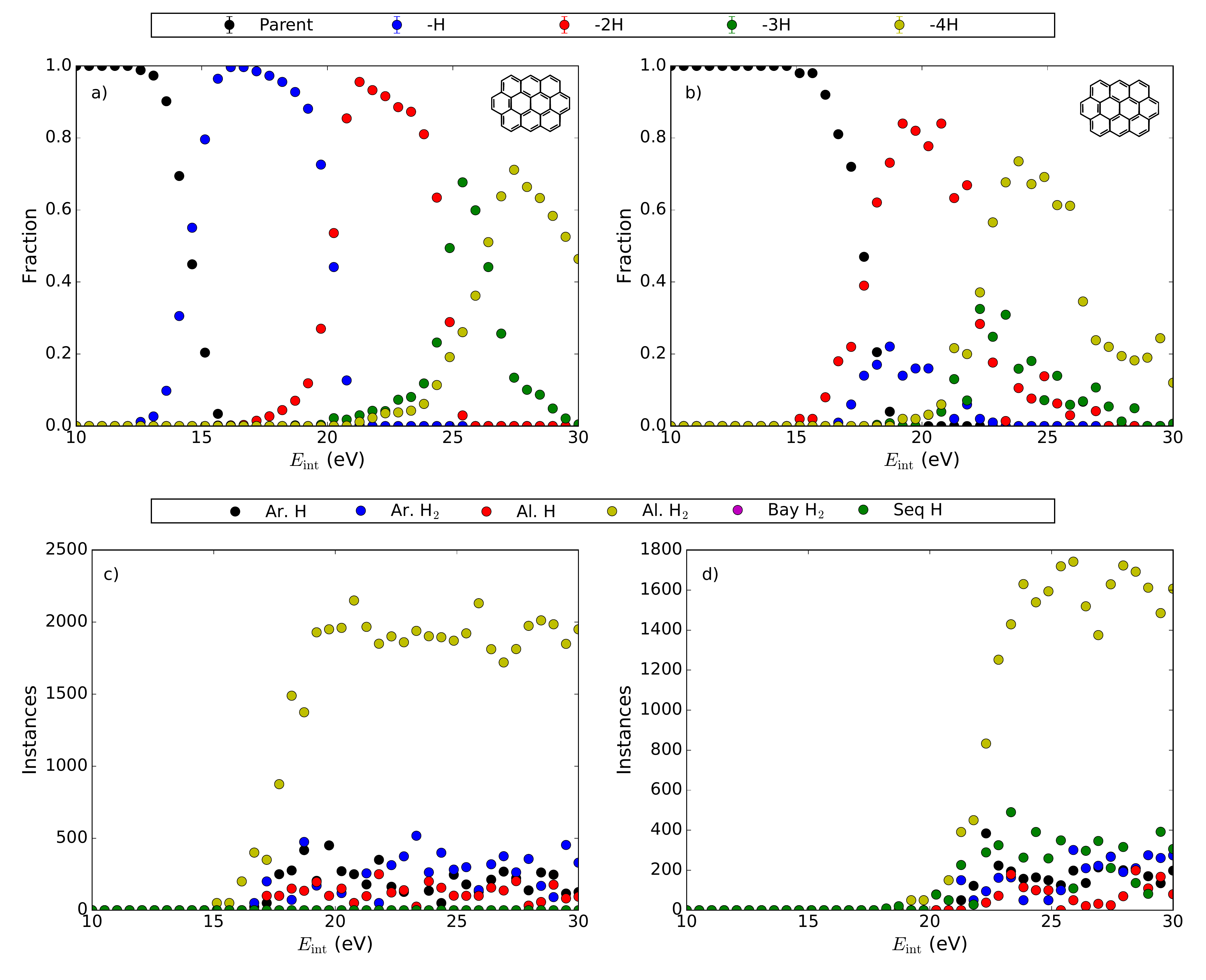}
  \caption{Results of Monte-Carlo simulations for ovalene. Panel (a) shows the simulation considering the barriers as per DFT calculations (including site variations; see Table~\ref{tab:trans_ova}), while panel (b) shows the results once the aliphatic loss barriers are set to the values from coronene. Panels (c) and (d) show the channel competition for the results shown in panel (b), with (c) showing the competition for the first fragmentation and (d) for the second fragmentation (from ovalene after \ce{H2}-loss).}
  \label{fig:Eint_ova}
\end{figure*}

Rearrangements of the carbon skeleton, including formation of ethynyl and vinyl groups, have recently been studied  \citep{bau14,bou16,tri17b} in  the context of PAH fragmentation. Since here we focus only on the dehydrogenation process, we calculated as an example the loss of H from coronene after the formation of an ethynyl group. This requires 5.41~eV, while loss of any other hydrogen  in the molecule needs only  4.87~eV. Thus we did not include these reaction paths in our simulations. Additionally, we tested the importance of ethynyl formation in coronene and found it to be negligible; while ethynyl groups are indeed formed, H-loss is not efficient in such configuration. The reverse reaction quickly will lead to the closure of the ring and dehydrogenation will occur from other structures, without significantly affecting the AEs (see Sect.~\ref{sec:mc_res} for more details). Thus, we have not considered isomerization channels involving the carbon skeleton.

\subsubsection{Results}
\label{sec:mc_res}

The Monte-Carlo simulations provide detailed insight in the main fragmentation processes but also highlight some discrepancies. While such discrepancies are minor for perylene and coronene, the case of ovalene must be further discussed before diving into the general results. When running the full Monte-Carlo simulations, it became immediately clear that hydrogen roaming to tertiary atoms posed an insurmountable challenge to the understanding of the experimental results. The barriers involved in H-shifts to tertiary atoms (see Table~\ref{tab:trans_ova}) imply that the solo sites act as attractors for H-atoms, leading to a fragmentation pattern that cannot be reconciled with our experiments. This issue led us to consider canceling hydrogen shifts to tertiary atoms. Such modification only has a minor effect in simulations involving perylene and coronene, but greatly modifies the case for ovalene. While it was necessary to implement further modifications to reproduce the experiments performed on ovalene (see later), canceling H-roaming to tertiary atoms remained indispensable. Given this, all results presented in this section refer to cases where H-roaming is limited to jumps within a single ring or across bay-regions. Simulation results involving the jumps to tertiary carbons are present in Appendix~\ref{sec:app}.

To validate our Monte-Carlo simulations we run them  under the conditions given by \citet{joc94} in an attempt to reproduce their results. In their experiments, they derive AEs of different fragmentation channels within a $10^{-4}$~s timescale for perylene and coronene, among other PAHs. For these two PAHs, they determined the AE of the H-loss (not observed in perylene and 12.05~eV for coronene) and \ce{H2}-loss channels (11.34 and 13.43~eV for perylene and coronene, respectively). Figure~\ref{fig:Eint}a and b show the results of the Monte-Carlo simulations up to 30~eV of internal energy for perylene and coronene; the vertical lines mark the AEs derived by \citet{joc94}. The energies at which the fragmentation products appear during the simulations show a reasonable agreement with the experiments by \citet{joc94}. The differences appear in the \ce{H2}/2H-loss channels; in perylene these become relevant at lower energy within our models with respect to the experimental values, while the opposite is true for coronene. The differences in both cases are  $\lesssim$1~eV and the non-detection of the H-loss channel in perylene is well reproduced as well, leading us to conclude that our Monte-Carlo simulation provides an acceptable model to study PAH fragmentation.

Additional channels (3H-loss and 4H-loss) are also observed at higher energies, with perylene undergoing quick full dehydrogenation. As was explained in Sect.~\ref{sec:theo}, we focus only in losses of up to four H-atoms since beyond this point other isomerization and fragmentation channels involving the carbon skeleton can become important. In these simulations, we have also investigated the different contributions to the fragmentation channels involved along the dehydrogenation process. Figure~\ref{fig:Eint}c and d clearly show that the first loss (\ce{H2} for perylene and H for coronene) occurs preferentially from aliphatic sites rather than from aromatic hydrogens. This means that hydrogen shifts play a significant role in PAH dehydrogenation. However, whether the fragmentation proceeds through H- or \ce{H2}-loss is molecule dependent. Figure~\ref{fig:Eint}c additionally shows that \ce{H2}-loss across the bay regions becomes a competing channel at high internal energies in perylene. The second loss in coronene (Fig.~\ref{fig:Eint}f) happens almost exclusively through the sequential hydrogen loss -- i.e.\ the lone hydrogen in the ring that has already lost a hydrogen atom. The same fragmentation channel dominates the second loss in perylene (Fig.~\ref{fig:Eint}e) in the beginning, due to the fact that a hydrogen atom crossing the bay region into an empty site is a very fast, almost barrierless reaction (Table~\ref{tab:trans_per}). This leads to a significant fraction of isomers with two H-atoms in two trios, leading in turn to four potential hydrogen losses with a reduced barrier. However, as the internal energy increases, the competition with the formation of a second aliphatic group and the loss of another \ce{H2} unit becomes relevant and both become equally likely.

We next applied the Monte-Carlo simulations to study PAH behavior under the conditions of our own experiments with i-PoP. The internal energy is now able to increase as the molecule absorbs photons during the simulated laser pulses. Figure~\ref{fig:ipop_sim}a and b shows the results for both perylene and coronene, respectively. While differences remain, the qualitative behavior of the simulations provides a good match to the differences observed experimentally (Fig.~\ref{fig:medge}b and c). In the case of perylene, we observe that the ratio of odd and even H-losses does increase as dehydrogenation progresses. However, the 3H:4H ratio derived experimentally is not as large at high laser fluence as the one obtained from the simulations. Coronene provides a closer match to the experimental data, with the H:2H and 3H:4H ratios reaching roughly the same values, although the fluence values at which the channels appear do not match exactly.

AEs of the dehydrogenation channels for ovalene have not been determined experimentally. Still, we looked into the fragmentation pattern and how well the simulation reproduces our own experiments.  As can be seen in Table~\ref{tab:trans_ova}, the various aliphatic losses from the duos have all different activation energies and enthalpies, depending on the position of the aliphatic groups. Figure~\ref{fig:Eint_ova}a shows the result of the simulation in terms of the internal energy, taking into consideration all these differences in the aliphatic losses. Comparison with the same simulation but for coronene (Fig.\ref{fig:Eint}a) suggests that this behavior with internal energy will not reproduce the experimental results. We have tested a set of possible solutions, considering all the aliphatic losses with the same parameters, whether they correspond to localized, delocalized or to the far-away carbons, but to no avail. In all these cases, the fragmentation pattern leads to different ratios of odd to even losses as dehydrogenation progresses or to the ratio being about twice as large as the experimental results. Due to the similarity of edge structure in coronene and ovalene, we explored  the possibility to set the energy barriers for aliphatic formation and losses in ovalene as in  coronene, while keeping the enthalpy of formation corresponding to  the delocalized aliphatic intermediate state of ovalene. The result is shown in Fig.~\ref{fig:Eint_ova}b; the fragmentation is now occurring in steps of two, with a small fraction of the molecules retaining 13 or 11 hydrogen atoms (1H or 3H loss). Given that this setting reproduces more closely the experimental results, we will present the results of  this simulation, and discuss the implications of our choice in the following section.

\begin{figure}[t!]
  \centering
  \includegraphics[width=\columnwidth]{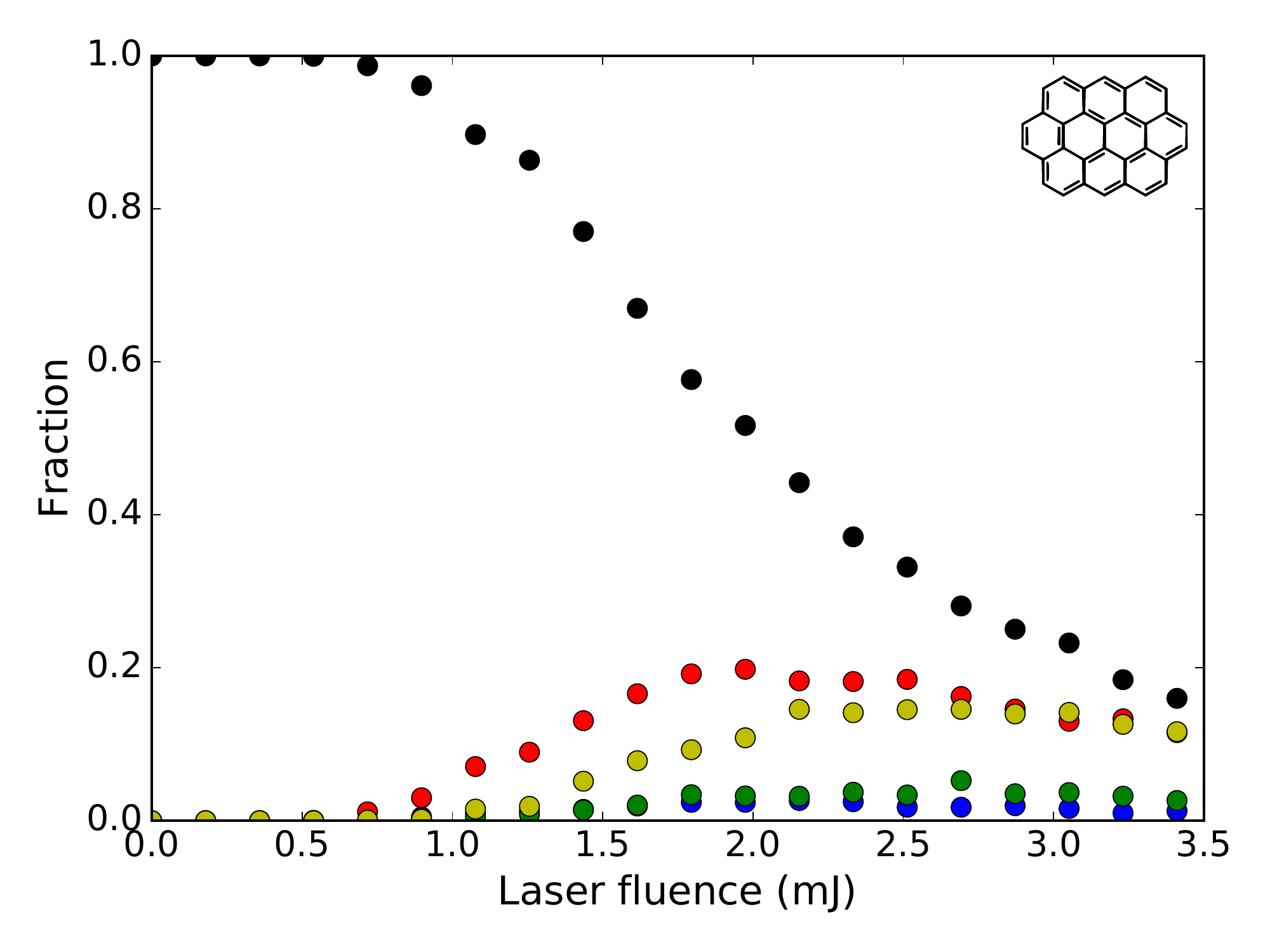}
  \caption{Results of Monte-Carlo simulations of the i-PoP experiments on the fragmentation of ovalene after irradiation with three laser pulses at 656~nm, following the simulation parameters used in Fig.~\ref{fig:Eint_ova}b. The colors follow the same scheme as those of the top panels of Fig.~\ref{fig:Eint_ova}.}
  \label{fig:ipop_ova}
\end{figure}

In ovalene, as in perylene, the dominant loss channel is aliphatic \ce{H2}-loss (Fig.~\ref{fig:Eint_ova}c), although a variety of other channels also contribute to a smaller extent. However, unlike perylene, ovalene has a compact structure, without bay-regions, which prevents the exchange of hydrogens between different rings after the first loss. This leads to the second loss (from ovalene after \ce{H2}-loss) being nearly identical in nature to the first one (Fig.~\ref{fig:Eint_ova}d). The results of the Monte-Carlo simulation considering the conditions of the experiment is shown in Fig.~\ref{fig:ipop_ova}. A qualitative comparison with the experimental results (Fig.~\ref{fig:solo}a), shows a remarkably similar behavior, with ovalene missing two hydrogens dominating at low laser fluence, and later ovalene missing four hydrogens reaching a similar level. Along the same lines, the low fraction of fragments with odd hydrogenation is very well reproduced, with both monodehydrogenated ovalene and ovalene with three lost hydrogens having nearly the same fractional abundance. However, it must be noted that the laser fluence range where the fragmentation happens in the Monte-Carlo simulations is much lower than that observed in the experiments.

\section{Discussion}
\label{sec:discussion}

The experimental results show that the edge structure  affects the intensity of successive odd and even hydrogenation losses in small PAHs, producing variation in the even-to-odd peak ratios as dehydrogenation progresses. Larger ($N_\mathrm{C} \geq 30$) PAHs behave similarly and show a pronounced decrease in the intensity of the odd peaks, with little to no variations in the odd to even ratios as the dehydrogenation progresses. The edge effect in small PAHs can be understood by directly looking at the energetics of the DFT calculations and is further supported by the results of our Monte-Carlo simulations. In all cases, the formation of aliphatic \ce{CH2} groups during hydrogen hopping is a crucial step in the dehydrogenation, since the vast majority of the losses from PAHs with an even number of hydrogens occur through aliphatic H- or \ce{H2}-loss. The preferred channel among these two appears to be highly dependent on the particular molecule. While the barriers for both processes tend to be close in energy (usually within the accuracy of DFT calculations), $\Delta S_{1000}$ shows larger variations.

The Monte-Carlo simulations for perylene and coronene allow us to understand the effect of edge structure and the different experimental dehydrogenation patterns both molecules exhibit. The absence of the -H fragment in perylene is due to the dominant channel being \ce{H2}-loss from an aliphatic site, while the increased intensity of the -3H fragment with respect to the -4H is explained as the rapid exchange of hydrogens between nearby rings across the bay regions.  On the other hand, coronene retains a considerable fraction of fragments with an odd number of hydrogens, although the evens are always more abundant, confirming earlier experiments \citep{joc94,job04}. Our simulations show that coronene loses hydrogens sequentially, thus explaining the pattern and the constant odd to even ratios. Aliphatic H-loss dominates the first loss, while the second hydrogen loss proceeds much more rapidly due to the reduced barrier in partially hydrogenated rings.

Using the results for coronene and perylene, we can rationalize the fragmentation patterns observed in the other molecules studied here and look for general trends applicable to PAHs as a whole. In Sect.~\ref{sec:theo_res} we saw that BDEs for H losses in duo and quarto behave similarly thus making the case for the overall fragmentation pattern --  enhanced even peaks over odd peaks -- observed in triphenylene. Furthermore, the presence of bay-regions and the possibility of a fast H hopping across them in this molecule also explains the differences observed between triphenylene and coronene, that is, odd peaks are slightly more intense in the former. The same case can be made for rubicene, where the presence of trios and bay regions in combination with quarto hydrogens gives raise to its unusual fragmentation pattern. All small PAHs ($N_\mathrm{C} < 32$) in our sample show variations in the dehydrogenation behavior that can be connected to the presence of bay-regions and the different edge structures present. The actual strength of odd-to-even ratios are indicative of whether H- or \ce{H2}-loss dominates the fragmentation. In our case, the low intensity for the first H-loss in perylene suggests that this is the only molecule which undergoes significant molecular hydrogen formation as part of its fragmentation process, a conclusion also supported by other studies \citep{joc94}.

Figure~\ref{fig:size} shows that the average odd-to-even ratio correlates with molecular size. Furthermore, the dependence on the edge structure appears to be washed away as the molecular size increases. This is particularly striking in the case of HBC: no variations in odd-to-even ratio are observed as dehydrogenation progresses, although the structure of HBC resembles that of perylene (trios and bay region), where the ratios do change. Such changes have been observed in other experiments involving HBC, although only when the hydrogenation level is much lower than what we currently investigated and with an overall lower degree of variations \citep{zhe14a}. The difference between perylene and HBC is probably due to a larger degree of competition between aliphatic and bay-region \ce{H2}-loss. This being the case, the influence of the H-hopping across bay regions would be greatly reduced and sequential H-loss would become a secondary channel until further down the dehydrogenation process.

The inability of our Monte-Carlo simulations to reproduce the experimental results for ovalene when all the possible rates are considered, tells us that there must be issues with the DFT results.  Given the structural similarity between coronene and ovalene (both compact molecules dominated by duo edges), it is tempting to speculate that the problem arises when dealing with the only structural difference, namely the presence of solo hydrogens. Hydrogens in solo positions are known to be the most reactive sites in PAHs \citep{aihara96}. Our quantum chemistry calculations show that once a H atom has moved from a duo group to the tertiary C between  solo and  duo (R1 from hereafter), the intermediate state lies quite high in energy (3.7~eV) with respect to the standard ovalene. From there, moving to the solo site (R2) requires only 0.17~eV --- almost a barrierless reaction --- which makes the solo site the place where the roaming H will end up being. From there, aliphatic H-loss has a lower barrier than \ce{H2}-loss and thus dominates. This situation would change if a) the barrier of R2 was higher or b) once on the tertiary C, the H would jump into the carbon skeleton rather than of along the edges. We then recalculated the barrier of R2 using the M06-2X functional, which is known to give more reliable values for barrier height \citep{m062x}, but we found a value 0.21~eV which does not affect the reaction rate. We also calculated the barrier for H-roaming inside the ring, and found that both forward and reverse barriers are comparable to other jumps to tertiary.

Only by using the barriers derived for coronene in the aliphatic formation and loss processes, we were able to produce a qualitative agreement with the experiments for ovalene. In order to explain the observed decrease of the odd-to-even ratio as PAH size increases (Fig.~\ref{fig:size}), the photofragmentation needs to be dominated by \ce{H2}-loss, whether from aliphatic sites or bay-regions. By looking in particular at the case of HBC, it is possible that the \ce{H2}-loss channel in bay-regions will become a more competitive channel --- as seen at high $E_\mathrm{int}$ in perylene. Aromatic \ce{H2}-loss, on the other hand, is never observed to be a competitive channel. This is expected due to its high energy barrier (more than 1~eV above most other dissociation channels). A summary of the dominant reactions according to the type of edge structure and PAH size is presented in Table~\ref{tab:react}.
 
\section{Conclusions}
\label{sec:conclusions}

In this paper we combined experiments, theory and modeling to understand the dehydrogenation behavior, in particular the odd-to-even ratio, of a sample of nine different and representative PAHs up to astronomical sizes $N_\mathrm{C}=48$. The results of the modeling are not only consistent to our own experimental results, but also reproduce reasonably well the observed appearance energies and fragmentation channels observed by \citet{joc94}. The Monte-Carlo simulations show two main factors acting on the behavior of the odd-to-even H-loss pattern observed during PAH dehydrogenation: edge structure and size. The edge structure effects are important in small PAHs and are related to the presence of trios and bay-regions,both of which increase the relative intensity of odd H peaks with respect to even peaks as dehydrogenation progresses. As PAH size increases, the odd-to-even peak ratio decreases, effectively reducing the variations due to different edge structures until they become almost unrecognizable for $N_{\mathrm{C}} \geq 32$). The decrease of the odd-to-even ratio with increasing PAH size can clearly be observed in Fig.~\ref{fig:size}, with the largest molecules displaying ratios close to zero. In line with the Monte-Carlo simulations, the lack of odd H peaks can be interpreted considering that in larger molecules aliphatic \ce{H2}-loss dominates over aliphatic H-loss.

\begin{table}[!t]
  \begin{adjustbox}{max width=\columnwidth}
    \begin{threeparttable}
      \caption{Summary of dehydrogenation reactions for PAHs.}
      \label{tab:react}
      \begin{tabular}{lcccc}
        \toprule
        Loss channel & \multicolumn{2}{c}{$N_\mathrm{C} < 32$} & \multicolumn{2}{c}{$N_\mathrm{C} \geq 32$} \\
        & Trio/Bay & Duo/Quarto & Trio/Bay & Duo/Quarto \\
        \midrule
        Aromatic H-loss & {\color{red}\checkmark} & {\color{red}\checkmark} & & \\
        Aromatic \ce{H2}-loss &  &  &  &  \\
        Aliphatic H-loss &  & \checkmark &  & \\
        Aliphatic \ce{H2}-loss & \checkmark &  & \checkmark{\color{red}\checkmark} & \checkmark{\color{red}\checkmark} \\
        Bay-region \ce{H2}-loss &  &  & \checkmark{\color{red}\checkmark} & \\
        \bottomrule
      \end{tabular}
      \begin{tablenotes}
      \item Black and red tick marks correspond to the first and second loss channel, respectively.
      \item Solo hydrogens can only be lost via aromatic H-loss.
      \end{tablenotes}
    \end{threeparttable}
  \end{adjustbox}
\end{table}

The loss channels from PAHs with even hydrogenation are all dominated exclusively by aliphatic losses, whether in the form of atomic or molecular hydrogen. Thus, hydrogen hopping along the edge of PAHs is an indispensable ingredient to the dehydrogenation of PAHs and should be taken into account when modelling the photochemistry of PAHs in space. This is particularly true when considering \ce{H2} formation on PAHs, given that such a loss from a fully aromatic molecule has a rate limiting barrier at much higher energy than the equivalent process from an aliphatic site.

H-hopping across different rings through a tertiary carbon shows little to no importance in small molecules and was problematic in the case of ovalene, resulting in ratios and fragmentation patterns inconsistent with the experimental results. This begs for a detailed assessment on the capabilities of DFT for studying fragmentation and isomerization processes in large molecules in general and solo containing PAHs in particular.

Our study could not elucidate which properties determine the preference for one loss channel or another. The size dependence observed in our experiments suggests that molecular size might be a crucial parameter in favoring \ce{H2}-loss. Further experiments, particularly in large PAHs with a significant number of solo hydrogens are needed to help elucidate the role of this edge structure in the dehydrogenation process and establish the accuracy of the rates and barriers derived from DFT calculations.

From an astrochemical perspective, this work suggests that molecular hydrogen can be an important by-product of the photodestruction of large, astronomically relevant PAHs. Such connection has been suggested before, but the vast majority of previous studies focus on hydrogen abstraction, which is only relevant in regions where PAHs are found in superhydrogenated states \citep[e.g., where $G_0/n(\mathrm{H})$ is very small, where $G_0$ is the UV field intensity and $n(\mathrm{H})$ the atomic hydrogen density:][]{Mont13,Bosch15,andrews16}. However, in regions where the UV radiation field is low, most of the H is typically already molecular. The current results point to the possibility of \ce{H2} formation at the very edge of PDRs (high $G_0/n(\mathrm{H})$), which is precisely where the transition between molecular and atomic hydrogen occurs. Astronomical models for PAH fragmentation based upon these results will help elucidate whether the fragmentation process can indeed act as an efficient factory for molecular hydrogen.

\begin{acknowledgements}
We thank M.J.A.\ Witlox and R.\ Koehler for technical support on i-PoP. Studies of interstellar chemistry at Leiden Observatory are supported through advanced-ERC grant 246976 from the European Research Council, through a grant by the Netherlands Organization for Scientific Research (NWO) as part of the Dutch Astrochemistry Network, and a Spinoza premie. We acknowledge the European Union (EU) and Horizon 2020 funding awarded under the Marie Sk\l{}odowska-Curie action to the EUROPAH consortium, grant number 722346. AC acknowledges NWO for a VENI grant (number 639.041.543). JZ acknowledges financial support from the Fundamental Research Funds for the Central Universities and from the National Science Foundation of China (NSFC, grant number 11743004). HL acknowledges NWO for a VICI grant (number 639.043.905). DFT calculations were carried out on the Dutch national e-infrastructure (Cartesius) with the support of SURF Cooperative, under NWO EW projects MP-270-13 and SH-362-15.
\end{acknowledgements}

\bibliographystyle{aa}
\bibliography{references}

\begin{thebibliography}{44}
\expandafter\ifx\csname natexlab\endcsname\relax\def\natexlab#1{#1}\fi

\bibitem[{Aihara {et~al.}(1996)Aihara, Fujiwara, Harada, Ichikawa, Fukushima,
  Hirota, \& Ishida}]{aihara96}
Aihara, J., Fujiwara, K., Harada, A., {et~al.} 1996, Theochem. J. Mol. Struct.,
  366, 219

\bibitem[{Andrews {et~al.}(2016)Andrews, Candian, \& Tielens}]{andrews16}
Andrews, H., Candian, A., \& Tielens, A. G. G.~M. 2016, \aap, 595, A23

\bibitem[{Baboul {et~al.}(2000)Baboul, Curtiss, Redfern, \&
  Raghavachari}]{bab00}
Baboul, A.~G., Curtiss, L.~A., Redfern, P.~C., \& Raghavachari, K. 2000, J.
  Chem. Phys., 110, 7650

\bibitem[{Baer \& Hase(1996)}]{baer-hase}
Baer, T. \& Hase, W.~L. 1996, Unimolecular reaction dynamics: theory and
  experiments No.~31 (Oxford University Press on Demand)

\bibitem[{Barker(2001)}]{Multiwell2}
Barker, J.~R. 2001, Int. J. Chem. Kinet., 33, 232

\bibitem[{Barker {et~al.}(2017)Barker, Nguyen, Stanton, Aieta, Ceotto, Gabas,
  Kumar, Li, Lohr, Maranzana, Ortiz, Preses, Simmie, Sonk, \&
  Stimac}]{Multiwell1}
Barker, J.~R., Nguyen, T.~L., Stanton, J.~F., {et~al.} 2017, MultiWell-2017
  Software Suite, \url{http://clasp-research.engin.umich.edu/multiwell/}, {J}.
  R. Barker, University of Michigan, Ann Arbor, Michigan, USA, 2017

\bibitem[{Bauschlicher(1998)}]{bau98}
Bauschlicher, C.~W., J. 1998, \apjl, 509, L125

\bibitem[{Bauschlicher \& Ricca(2014)}]{bau14}
Bauschlicher, C.~W., J. \& Ricca, A. 2014, Theor. Chem. Acc., 133, 1479

\bibitem[{Bern\'e {et~al.}(2015)Bern\'e, Montillaud, \& Joblin}]{ber15}
Bern\'e, O., Montillaud, J., \& Joblin, C. 2015, \aap, 577, A133

\bibitem[{Bern\'e \& Tielens(2012)}]{ber12}
Bern\'e, O. \& Tielens, A. G.~G.~M. 2012, Proc. Natl. Acad. Sci., 109, 401

\bibitem[{Boersma {et~al.}(2012)Boersma, Rubin, \& Allamandola}]{boe12}
Boersma, C., Rubin, R.~H., \& Allamandola, L.~J. 2012, \apj, 753, 168

\bibitem[{Boschman {et~al.}(2015)Boschman, Cazaux, Spaans, Hoekstra, \&
  Schlath{\"o}lter}]{Bosch15}
Boschman, L., Cazaux, S., Spaans, M., Hoekstra, R., \& Schlath{\"o}lter, T.
  2015, \aap, 579, A72

\bibitem[{Bouwman {et~al.}(2016)Bouwman, de~Haas, \& Oomens}]{bou16}
Bouwman, J., de~Haas, A.~J., \& Oomens, J. 2016, Chem. Commun., 52, 2636

\bibitem[{Cami {et~al.}(2010)Cami, Bernard-Salas, Peeters, \& Malek}]{Cami2010}
Cami, J., Bernard-Salas, J., Peeters, E., \& Malek, S.~E. 2010, Science, 329,
  1180

\bibitem[{Castellanos {et~al.}(2014)Castellanos, Bern\'e, Sheffer, Wolfire, \&
  Tielens}]{cas14}
Castellanos, P., Bern\'e, O., Sheffer, Y., Wolfire, M.~G., \& Tielens, A.
  G.~G.~M. 2014, \apj, 794, 83

\bibitem[{Chen {et~al.}(2015)Chen, Gatchell, Stockett, Delaunay, Domaracka,
  Micelotta, Tielens, Rousseau, Adoui, Huber, Schmidt, Cederquist, \&
  Zettergren}]{Chen15}
Chen, T., Gatchell, M., Stockett, M.~H., {et~al.} 2015, J. Chem. Phys., 142,
  144305

\bibitem[{Dyakov {et~al.}(2006)Dyakov, Ni, Lin, Lee, \& M.}]{dya06}
Dyakov, Y.~A., Ni, C.-K., Lin, S.~H., Lee, Y.~T., \& M., M.~A. 2006, Phys.
  Chem. Chem. Phys., 8, 1404

\bibitem[{Ekern {et~al.}(1997)Ekern, Marshall, Szczepanski, \& Vala}]{eke97}
Ekern, S.~P., Marshall, A.~G., Szczepanski, J., \& Vala, M. 1997, \apjl, 488,
  L39

\bibitem[{Ekern {et~al.}(1998)Ekern, Marshall, Szczepanski, \& Vala}]{eke98}
Ekern, S.~P., Marshall, A.~G., Szczepanski, J., \& Vala, M. 1998, J. Phys.
  Chem. A, 102, 3498

\bibitem[{Frisch {et~al.}(2009)Frisch, Trucks, Schlegel, Scuseria, Robb,
  Cheeseman, Scalmani, Barone, Mennucci, Petersson, Nakatsuji, Caricato, Li,
  Hratchian, Izmaylov, Bloino, Zheng, Sonnenberg, Hada, Ehara, Toyota, Fukuda,
  Hasegawa, Ishida, Nakajima, Honda, Kitao, Nakai, Vreven, Montgomery, Peralta,
  Ogliaro, Bearpark, Heyd, Brothers, Kudin, Staroverov, Kobayashi, Normand,
  Raghavachari, Rendell, Burant, Iyengar, Tomasi, Cossi, Rega, Millam, Klene,
  Knox, Cross, Bakken, Adamo, Jaramillo, Gomperts, Stratmann, Yazyev, Austin,
  Cammi, Pomelli, Ochterski, Martin, Morokuma, Zakrzewski, Voth, Salvador,
  Dannenberg, Dapprich, Daniels, Farkas, Foresman, Ortiz, Cioslowski, \&
  Fox}]{fri09}
Frisch, M.~J., Trucks, G.~W., Schlegel, H.~B., {et~al.} 2009, Gaussian 09
  {R}evision {D}.01, {G}aussian Inc. Wallingford CT 2009

\bibitem[{Garc{\'i}a-Hern{\'a}ndez {et~al.}(2012)Garc{\'i}a-Hern{\'a}ndez,
  Villaver, Garc{\'i}a-Lario, Acosta-Pulido, Manchado, Stanghellini, Shaw, \&
  Cataldo}]{gar12}
Garc{\'i}a-Hern{\'a}ndez, D.~A., Villaver, E., Garc{\'i}a-Lario, P., {et~al.}
  2012, \apj, 760, 107

\bibitem[{Joblin(2003)}]{job03}
Joblin, C. 2003, in SF2A-2003: Semaine de l'Astrophysique Fran\c{c}ais, ed.
  F.~Combes, D.~Barret, T.~Contini, \& L.~Pagani (Les Ulis, France: EDP
  Sciences), 175

\bibitem[{Joblin(2004)}]{job04}
Joblin, C. 2004, in The Dense Interstellar Medium in Galaxies, ed. S.~Pfalzner,
  C.~Kramer, C.~Staubmeier, \& A.~Heithausen, Vol.~91, 517

\bibitem[{Jochims {et~al.}(1994)Jochims, R{\"u}hl, Baumg{\"a}rtel, Tobita, \&
  Leach}]{joc94}
Jochims, H.~W., R{\"u}hl, E., Baumg{\"a}rtel, H., Tobita, S., \& Leach, S.
  1994, \apj, 420, 307

\bibitem[{Ling {et~al.}(1995)Ling, Gotkis, \& Lifshitz}]{Ling95}
Ling, Y., Gotkis, Y., \& Lifshitz, C. 1995, Eur. J. Mass Spectrom., 1, 41

\bibitem[{Ling \& Lifshitz(1998)}]{Ling98}
Ling, Y. \& Lifshitz, C. 1998, J. Phys. Chem. A, 102, 708

\bibitem[{Malloci {et~al.}(2007)Malloci, Joblin, \& Mulas}]{mal07}
Malloci, G., Joblin, C., \& Mulas, G. 2007, \aap, 462, 627

\bibitem[{Montgomery {et~al.}(2000)Montgomery, Frisch, \& Ochterski}]{mon00}
Montgomery, J.~A., J., Frisch, M.~J., \& Ochterski, J.~W. 2000, J. Chem. Phys.,
  112, 6532

\bibitem[{Montillaud {et~al.}(2013)Montillaud, Joblin, \& Toublanc}]{Mont13}
Montillaud, J., Joblin, C., \& Toublanc, D. 2013, \aap, 552, A15

\bibitem[{Otsuka {et~al.}(2013)Otsuka, Kemper, Hyung, Sargent, Meixner,
  Tajitsu, \& Yanagisawa}]{ots13}
Otsuka, M., Kemper, F., Hyung, S., {et~al.} 2013, \apj, 764, 77

\bibitem[{Paris {et~al.}(2014)Paris, Alcam{\'i}, Mart{\'i}n, \&
  D{\'i}az-Tendero}]{Paris14}
Paris, C., Alcam{\'i}, M., Mart{\'i}n, F., \& D{\'i}az-Tendero, S. 2014, J.
  Chem. Phys., 140, 204307

\bibitem[{Peng {et~al.}(1996)Peng, Ayala, Schlegel, \& Frisch}]{STQN2}
Peng, C., Ayala, P.~Y., Schlegel, H.~B., \& Frisch, M.~J. 1996, J. Comput.
  Chem., 17, 49

\bibitem[{Peng \& Schlegel(1993)}]{STQN1}
Peng, C. \& Schlegel, H.~B. 1993, Isr. J. Chem., 33, 449

\bibitem[{Reed \& Kass(2000)}]{ree00}
Reed, D.~R. \& Kass, S.~R. 2000, J. Mass Spectrom., 35, 534

\bibitem[{Rodriguez~Castillo {et~al.}(2018)Rodriguez~Castillo, Simon, \&
  Joblin}]{cas18}
Rodriguez~Castillo, S., Simon, A., \& Joblin, C. 2018, Int. J. Mass Spectrom.

\bibitem[{Sellgren {et~al.}(2010)Sellgren, Werner, Ingalls, Smith, Carleton, \&
  Joblin}]{sell10}
Sellgren, K., Werner, M.~W., Ingalls, J.~G., {et~al.} 2010, \apjl, 722, L54

\bibitem[{Tielens(2013)}]{Tielens2013}
Tielens, A. G. G.~M. 2013, Rev. Mod. Phys., 85, 1021

\bibitem[{Trinquier {et~al.}(2017{\natexlab{a}})Trinquier, Simon, Rapacioli, \&
  Gad{\'e}a}]{tri17a}
Trinquier, G., Simon, A., Rapacioli, M., \& Gad{\'e}a, F.~X.
  2017{\natexlab{a}}, Mol. Astrophys., 7, 27

\bibitem[{Trinquier {et~al.}(2017{\natexlab{b}})Trinquier, Simon, Rapacioli, \&
  Gad{\'e}a}]{tri17b}
Trinquier, G., Simon, A., Rapacioli, M., \& Gad{\'e}a, F.~X.
  2017{\natexlab{b}}, Mol. Astrophys., 7, 37

\bibitem[{Wakelam {et~al.}(2017)Wakelam, Bron, Cazaux, Dulieu, Gry, Guillard,
  Habart, Hornek{\ae}r, Morisset, Nyman, Pirronello, Price, Valdivia, Vidali,
  \& Watanabe}]{Wakelam2017}
Wakelam, V., Bron, E., Cazaux, S., {et~al.} 2017, Mol. Astrophys., 9, 1

\bibitem[{West {et~al.}(2014)West, Useli-Bacchitta, Sabbah, Blanchet, Bodi,
  Mayer, \& Joblin}]{job14}
West, B., Useli-Bacchitta, F., Sabbah, H., {et~al.} 2014, J. Phys. Chem. A,
  118, 7824

\bibitem[{Zhao \& Truhlar(2011)}]{m062x}
Zhao, Y. \& Truhlar, D.~G. 2011, J. Chem. Theory and Comput., 7, 669

\bibitem[{Zhen {et~al.}(2014{\natexlab{a}})Zhen, Castellanos, Paardekooper,
  Linnartz, \& Tielens}]{zhe14b}
Zhen, J., Castellanos, P., Paardekooper, D.~M., Linnartz, H., \& Tielens, A.
  G.~G.~M. 2014{\natexlab{a}}, \apjl, 797, L30

\bibitem[{Zhen {et~al.}(2014{\natexlab{b}})Zhen, Paardekooper, Candian,
  Linnartz, \& Tielens}]{zhe14a}
Zhen, J., Paardekooper, D.~M., Candian, A., Linnartz, H., \& Tielens, A.
  G.~G.~M. 2014{\natexlab{b}}, Chem. Phys. Lett., 592, 211

\end{thebibliography}

\begin{appendix}

\section{Monte-Carlo with tertiary H-hopping}
\label{sec:app}

\begin{figure}[ht!]
  \centering
  \includegraphics[width=\columnwidth]{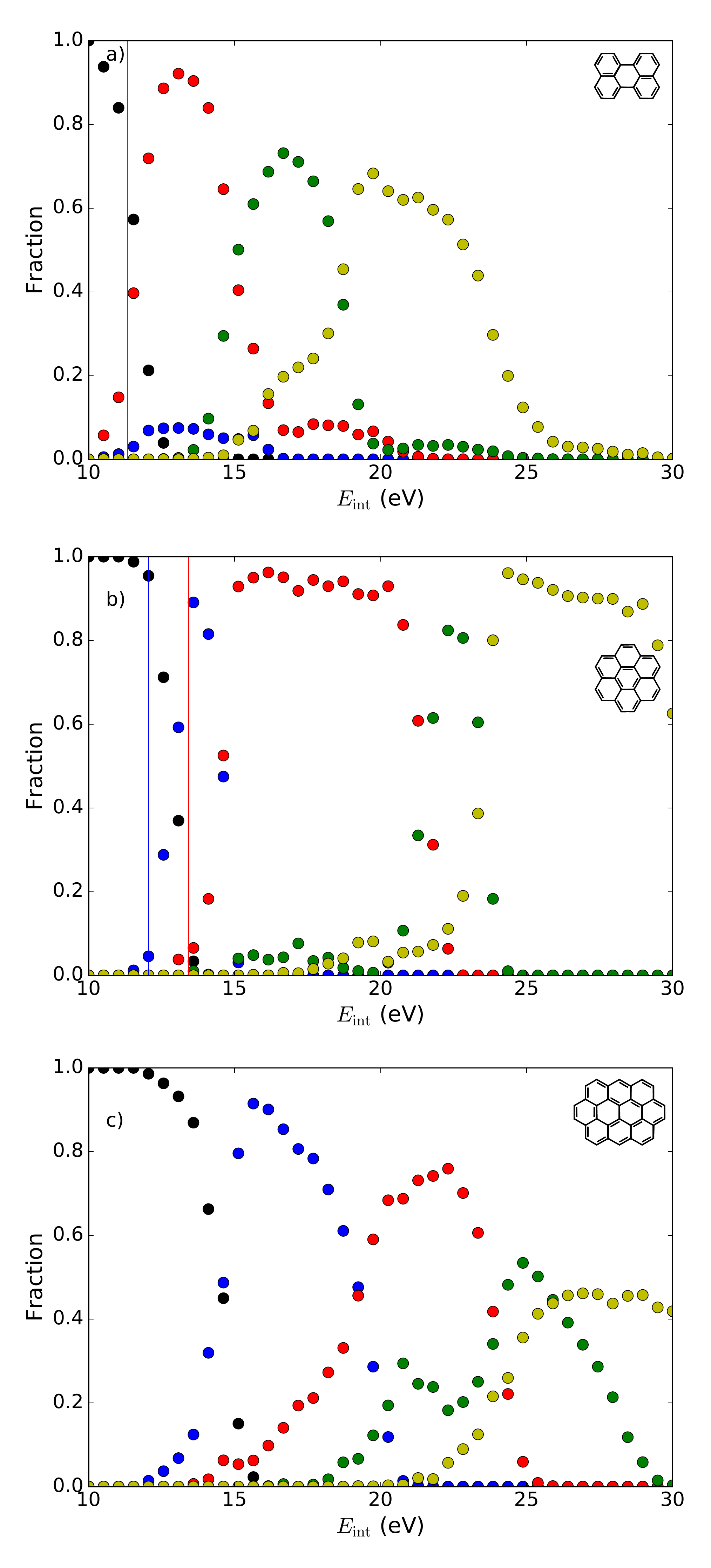}
  \caption{Monte-Carlo results for perylene, coronene and ovalene fragmentation as a function of internal energy. These results take into consideration the possibility of H-hopping involving tertiary carbons. The colors follow the same scheme as those of the top panels of Fig.~\ref{fig:Eint}.}
  \label{fig:Eint_tert}
\end{figure}

Figure~\ref{fig:Eint_tert} shows the results of the Monte-Carlo simulations of perylene, coronene and ovalene. In this case all rates have been allowed, including H-hopping to and from tertiary carbons, which were excluded in the main text. Perylene (Fig.~\ref{fig:Eint_tert}a) shows no noticeable changes with respect to the results with exclusion of the tertiary rates (Fig.~\ref{fig:Eint}a). Coronene (Fig.~\ref{fig:Eint_tert}b) displays minor changes in AE for its fragments with respect to the results from Fig.~\ref{fig:Eint}b, but the overall behavior is not modified. Note that in the present case the match with the AE of \citet{joc94} is better. However, the case of ovalene (Fig.~\ref{fig:Eint_tert}c) is markedly different from the simulation run with no tertiary jumps (Fig.~\ref{fig:Eint_ova}a and b). As explained in the main text, this is due to the solo hydrogens acting as attractors for the roaming hydrogens.

\end{appendix}

\end{document}